\documentclass[fleqn,usenatbib]{mnras}

\usepackage{newtxtext,newtxmath}

\usepackage[T1]{fontenc}

\DeclareRobustCommand{\VAN}[3]{#2}
\let\VANthebibliography\thebibliography
\def\thebibliography{\DeclareRobustCommand{\VAN}[3]{##3}\VANthebibliography}

\usepackage{graphicx}	
\usepackage{amsmath}	
\usepackage[flushleft]{threeparttable}
\usepackage{booktabs,caption,fixltx2e}
\usepackage{float}
\usepackage{comment}
\usepackage{tabularx}
\usepackage{subcaption}

\title[The effect of precession on light curves of TDEs]{The effect of relativistic precession on light curves of tidal disruption events}

\author[D. Calder\'on et al.]{
Diego Calder\'on,$^{1,2}$\thanks{E-mail: calderon@hs.uni-hamburg.de (DC)}
Ond\v{r}ej Pejcha,$^{1}$ Brian D. Metzger,$^{3,4}$~and Paul C. Duffell$^{5}$
\\
$^{1}$Institute of Theoretical Physics, Faculty of Mathematics and Physics, Charles University, 180 00 Prague, Czech Republic\\
$^{2}$Hamburger Sternwarte, Universit\"at Hamburg, Gojenbergsweg 112, 21029 Hamburg, Germany\\
$^{3}$Department of Physics and Columbia Astrophysics Laboratory, Columbia University, New York, NY 10027, USA\\
$^{4}$Center for Computational Astrophysics, Flatiron Institute, 162 5th Avenue, New York, NY 10010, USA\\
$^{5}$Department of Physics and Astronomy, Purdue University, 525 Northwestern Avenue, West Lafayette, IN 47907-2036, USA
}

\date{Accepted XXX. Received YYY; in original form ZZZ}

\pubyear{2024}

\begin{document}
\label{firstpage}
\pagerange{\pageref{firstpage}--\pageref{lastpage}}
\maketitle

\begin{abstract}
    The disruption of a star by the tidal forces of a spinning black hole causes the stellar stream to precess affecting the conditions for triggering the tidal disruption event (TDE). In this work, we study the effect that precession imprints on TDE light curves due to the interaction of the TDE wind and luminosity with the stream wrapped around the black hole. We perform two-dimensional radiation-hydrodynamic simulations using the moving-mesh hydrodynamic code JET with its radiation treatment module. We study the impact of black hole mass, accretion efficiency, and inclination between the orbital and spin planes. From our results, we identified two behaviours: $i)$ models with low-mass black holes ($M_\text{h}\sim10^6~\text{M}_{\odot}$), low inclination ($i\sim0$), and low accretion efficiency ($\eta\sim0.01$) show light curves with a short early peak caused by the interaction of the wind with the inner edge of the stream. The line of sight has little effect on the light curve, since the stream covers a small fraction of the solid angle due to the precession occurring in the orbital plane; $ii)$ models with high-mass black holes ($M_\text{h}\gtrsim10^7~\text{M}_{\odot}$), high inclination ($i\sim90^{\circ}$), and high accretion efficiency ($\eta\sim0.1$) produce light curves with luminosity peaks that can be delayed by up to 50-100 d depending on the line of sight due to presence of the precessed stream blocking the radiation in the early phase of the event. Our results show that black hole spin and misalignment do not imprint recognisable features on the light curves but rather can add complications to their analysis.
\end{abstract}



\section{Introduction}
\label{sec:intro}
    Most, if not all, galaxies in the Universe harbour a super-massive black hole at their centre \citep{ferrarese2005,kormendy2013}. 
    Yet, detecting them is challenging because they typically do not accrete at high enough rates to generate observable radiation. 
    This can change when a star is ripped apart by the tidal forces of the black hole, as this process produces a bright transient known as a tidal disruption event (TDE). 
    The outcome of this process is that half of the stellar material becomes unbound while the other half remains bound and returns to pericentre,  forming an accretion disc and producing a bright flare that can last for days to years. \citep{hills1975,rees1988,alexander2005}. 
    The light curve of the event is characterised by a fast rise to a luminosity peak of about 10$^\text{42}$-10$^\text{44}$~erg~s$^\text{-1}$ with a decay $\propto t^{-5/3}$ that follows the rate at which the mass returns to pericentre \citep[e.g.][]{gezari2006}. 
    Although if the star is not fully disrupted the power law of the decay may be different \citep{phinney1989,guillochon2013}.   
    A complete physical understanding of TDEs could provide a method for estimating the properties of dormant nuclei, i.e. the black hole mass and spin, where dynamical measurements are not possible. 
    In addition, TDEs can be used as laboratories to study accretion discs, jet formation, as well as gravitational wave emission \citep[e.g.][]{toscani2021,pfister2022,amaro-seoane2023}.

    The development of surveys for the detecting transients, such as iPTF \citep{rau2009,blagorodnova2019}, ASAS-SN \citep{kochanek2017,wevers2019,hinkle2021}, Pan-STARRS \citep{chambers2016,nicholl2019}, ZTF \citep{bellm2019,vanvelzen2020,vanvelzen2021}, SRG/eROSITA \citep{predehl2021,sazonov2021} has contributed to increase the sample of events with observations across most of the electomagnetic spectrum \citep[see][for a review]{gezari2021}. 
    To date, most TDEs have been detected in optical and ultraviolet wavelengths, or in X-ray emission. 
    But only a few events have been observed in both optical/UV and X-ray frequencies \citep{gezari2021}.
    \cite{dai2018} proposed that this property is due to a line-of-sight effect: the X-ray light can only escape along directions close to the poles, while the rest is reprocessed in an optically thick outflow. 
    However, despite the increase of the TDE sample there are crucial physical aspects that are not yet fully understood. 
    For instance, it is not clear what the power source is: a small amount of material accreted by the black hole \citep{metzger2016}, dissipation of the stream self-interaction \citep{piran2015,jiang2016}, and/or secondary shocks \cite{bonnerot2020}. 
    Additionally, some light curves display enigmatic features such as rebrightenings or plateaus \citep[e.g.][]{blanchard2017,godoy2017,wevers2019} that differ significantly from the general picture.
    
    The theoretical foundations of TDEs were developed by \cite{hills1975}, \cite{carter1983}, \cite{rees1988}, and \cite{phinney1989}, while the first numerical study was carriet out by \cite{evans1989}.
    Since then, numerical efforts have been focused on a wide variety of aspects of the evolution: the disruption of the star the subsequent evolution of the tidal stream \citep{lodato2009,guillochon2013,mainetti2017}, the quantification of the role of magnetic fields \citep{guillochon2017,bonnerot2017,curd2019}, the impact of using more realistic stellar structures \citep{goicovic2019,golightly2019,law-smith2019}, the formation of the disc from the stellar stream \citep{rosswog2009,hayasaki2013,bonnerot2016,bonnerot2020,metzger2022}, and even considering the relativistic effects due to the large mass of the black hole and its rotation \citep{kesden2012,gafton2019,liptai2019,ryu2023,jankovic2023a}, among others \citep[see][for a review]{lodato2020}. 

    The development of numerical tools and models that take account of the relativistic effects has helped to explain certain features observed in the light curves. 
    Relativistic precession due to the spin of the black hole can prevent the self-collision of the tidal stream on a single winding \citep{stone2012,dai2013,guillochon2015}. 
    Instead, the stream may wrap around the black hole several times before the event is triggered, delaying the generation of the electromagnetic signal. 
    Furthermore, \cite{guillochon2015} showed that massive black holes ($\gtrsim10^7~\text{M}_{\odot}$) can cause prompt TDE flares due to more violent self-collisions, which has been invoked to explain light curve signatures with rapid rises and plateaus \citep{wevers2019}. 
    This problem has been revisited by \cite{batra2023}, who solved the approximate tidal equation of the stream to calculate its thickness evolution during its trajectories along geodesics.
    However, there is a crucial aspect that has not been considered in this scenario: what happens to the remnant of the tidal stream wrapped around the black hole, once the event is triggered.
    In principle, once this occurs the luminosity and the disc wind will imminently encounter part of stream. 
    If this is the case, it is reasonable to ask whether such interactions can leave detectable features in the TDE light curves, as they could give us information about the spin of the black hole. 

    Semi-analytical models to describe how radiation is reprocessed through a dense wind have been used in the context of TDEs \citep[e.g.][]{metzger2016,kremer2023,uno2023c}. 
    For instance, \cite{piro2020} developed a formalism to estimate how much radiation escapes from a spherically symmetric outflow being illuminated by a central source. 
    However, these methods are limited by the spherical symmetry assumption. 
    To overcome this obstacle, \cite{calderon2021} performed radiation hydrodynamic simulations of wind-reprocessed transients, managing to validate the formalism of \cite{piro2020}. 
    Additionally, they demonstrated the capabilities of their numerical tool to model mixtures of gas-radiation multi-dimensionally and over wide ranges of space and time. 
    This highlights the relevance of using efficient and multi-dimensional numerical tools to model transient phenomena with complex geometries.
    
    In this work, we present a set of radiation-hydrodynamic simulations of the impact that relativistic precession can have on the light curves of TDEs. 
    Our model considers that the stellar stream wraps around the black hole until self-collision is achieved based on the approach by \cite{guillochon2015}. 
    We perform two-dimensional moving-mesh radiation-hydrodynamic simulations of the interaction of the TDE radiation and disc wind with the rest of the stream wrapped around the black hole. 
    The calculations were done using the moving-mesh hydrodynamic code JET \citep{duffell2011,duffell2013} with the radiation treatment module that we have developed \citep{calderon2021}.
    For rapidly spinning black holes, we find that such interactions can modify the expected light curves by up to $\sim$50\% during the first $100$-$200$ days of the event, depending on the line of sight and the black hole mass.
    Furthermore, under extreme conditions our model predicts either the appearance of an early peak of roughly of the same amplitude as the main maximum, or a delay in the peak of the expected light curve.
    This paper is structured as follows: Section~\ref{sec:model} describes the model and its assumptions, Section~\ref{sec:numerical} specifies the numerical method and setup for conducting the simulations, Section~\ref{sec:results} presents the results of the hydrodynamics and light curves computed from the simulations, Section~\ref{sec:discussion} details the observational implications of our results and limitations of our model. 
    Finally in Section~\ref{sec:conclusions} we summarise our findings and discuss further guidelines.
    
\section{The model}
\label{sec:model}
    \subsection{Analytical foundations}
    \label{sec:analytical}
\begin{figure}
            \centering
            \includegraphics[width=0.475\textwidth]{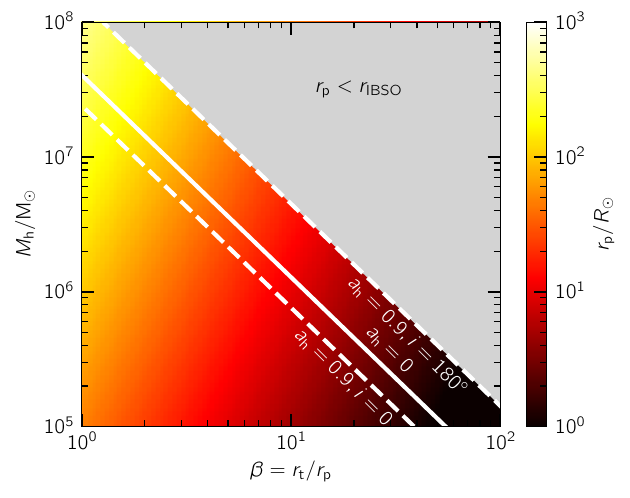}
            \caption{
	    Parameters of a tidal disruption event. 
            Pericentre distance $r_\text{p}$ as a function of black hole mass $M_\text{h}$ and penetration parameter $\beta$. 
            The region above the white lines represents the parameter space, where a star would enter the radius of the innermost bound spherical orbit $r_\text{IBSO}$. 
            The solid white line represent the case of a non-spinning black hole. 
            The dashed white lines correspond to the cases of a fast spinning black hole with $a_\text{h}=0.9$, and inclinations $i=0, 180^{\circ}$. 
            Tidal disruption events can take place only for parameters below the white lines.
            }
            \label{fig:ibso}
        \end{figure}

        Before presenting our model we introduce the quantities for describing a TDE. 
        First, let us consider a star of mass $m_*$ and radius $R_*$, and a black hole of mass $M_\text{h}$. 
        If the star travels inside the tidal radius $r_\text{t}$, i.e. the distance at which the black hole tidal force equates the stellar self-gravity, it will be disrupted. 
        The tidal radius is given by
	\begin{equation}
	    r_\text{t}=R_*q^{1/3},
        \end{equation}
        where $q=M_\text{h}/m_*$ is the mass ratio.
        It is important to remark that the star will be disrupted if it enters the sphere of radius $r_\text{t}$, as long as its pericentre is not smaller than the radius of the innermost bound spherical orbit $r_{\rm IBSO}$ \citep{bardeen1972} given by 
        \begin{equation}
           r_{\rm IBSO} = \frac{1}{2}R_{\rm Sch}\left(1+\sqrt{1\pm a_\text{h}}\right)^2,
	  \end{equation}
        where, $a_\text{h}$ is the spin parameter of the black hole, $R_\text{Sch}$ is its Schwarzschild radius, and the upper (``$+$") and lower (``$-$") signs corresponds to the the prograde ($i=0$) and retrograde ($i=180^{\circ}$) cases, respectively\footnote{To the best of our knowledge, there is no general analytical expression for $r_\text{IBSO}$ with $i\neq0,180^{\circ}$}.  
        Otherwise, the star will enter the black hole before the generation of electromagnetic radiation can take place. 
        Fig.~\ref{fig:ibso} shows the combinations of black hole mass $M_\text{h}$ and penetration ratio $\beta$ that allow a tidal disruption event to occur. 
        The penetration ratio is defined as $\beta=r_\text{t}/r_\text{p}$, where $r_\text{p}$ corresponds to the pericentre distance of the stellar orbit.
        The region above each white line highlights the parameters, where the star would plunge directly into the black hole, i.e. when the stellar pericentre is shorter than the innermost bound circular orbit.
        
        Once the star crosses the the tidal radius, it becomes an elongated stream that is being compressed along the axes that are perpendicular to its orbital velocity $\mathbf{v}$. 
        During pericentre passage the material is heated up, increasing its entropy and, as a consequence the width of the stream. 
        If there is precession out of the orbital plane the stream can go through pericentre passage multiple times before it collides with itself and triggers the event. 
        General relativistic effects can produce precession that change the orbital plane if the black hole has significant spin and there is misalignment between its rotation axis and the stream orbital plane. 
        Nevertheless, modelling this process is not straightforward, as calculating the trajectories and properties of the stream under the influence of a strong gravitational field implies the use of general relativistic hydrodynamic simulations.         
        
        Here, we follow the semi-analytical approach developed by \cite{guillochon2015} to estimate the width of the stellar stream as well as the total number of windings.
        The method assumes that the width of the stream $S$ is proportional to the ratio between the velocity perpendicular and parallel velocities to the orbital plane, i.e. $S\propto v_{\perp}/v_{\parallel}$. 
        Then, it considers that the growth of the perpendicular speed $v_{\perp}$ during each pericentre passage is proportional to the number of windings $W$ of the stellar stream around the black hole, i.e.
        \begin{equation}
            v_{\perp}=W\beta v_\text{esc},
            \label{eq:vperp}
        \end{equation}
        i.e. the spread in velocity increases in $\beta v_\text{esc}$ in every passage, being $v_\text{esc}$ the escape velocity of the star.  
        
        Now, we consider the stream as a cylinder with a radius given by
        \begin{equation}
            S=\frac{rv_{\perp}}{|\mathbf{v}|}\approx W\beta r q^{-1/3},
        \label{eq:radius}
        \end{equation}
        where we have used the equation~\ref{eq:vperp}, combined with $|\mathbf{v}|\sim\sqrt{2GM_\text{h}/r}$, $v_\text{esc}=\sqrt{2Gm_*/R_*}$. 
        This implicitly assumes that the self-gravity of the stream is not relevant after the first pericentre passage: $R_*=rq^{-1/3}$. 
        
        The equation~\ref{eq:radius} indicates that the width of the stream is proportional to the distance to the black hole and the number of times it has returned to pericentre. 
        In reality this is more complex, as part of the stream could be confined by self-gravity \citep{coughlin2016,steinberg2019}, and  cooling and recombination processes are expected to play a significant role \citep{kochanek1994,hayasaki2013,bonnerot2016,hayasaki2016,metzger2022}. 
        However, this approach still gives us a general description of the stream, which is useful due to its simple implementation for the generation of initial conditions in our simulations (see Section~\ref{sec:ics}). 
        Capturing the detailed structure of the stream is beyond the scope of this work, as our main goal is to quantify the impact of surrounding structures in the light curves of TDEs.
        
        \begin{figure*}
            \centering
	    \includegraphics[width=0.95\textwidth]{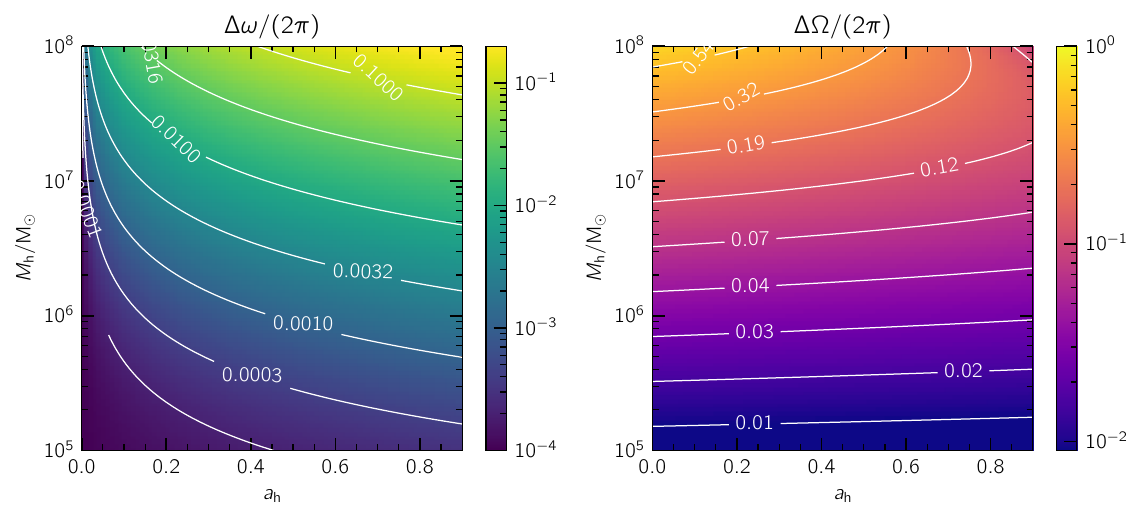}
	    \caption{
            Orbit-averaged nodal $\Delta\omega$ (left panel) and apsidal precession $\Delta\Omega$ (right panel) as a function of black hole mass $M_{\rm h}$ and spin parameter $a_\text{h}$ for events with penetration parameter $\beta=1$ and inclination between black hole rotation and orbital plane $i=45^{\circ}$ estimated using Equations~\ref{eq:press1}-\ref{eq:press7}.
            }
            \label{fig:precession}
        \end{figure*}
        
        To estimate the trajectory and how many times the stream wraps around the black hole we consider that the stream can be described as a set of disconnected ellipses. 
        Then, during every pericentre passage a new ellipse is generated but it is rotated according to the effects of relativistic precession calculated by the following post-Newtonian expressions \citep{merritt2013}
	\begin{eqnarray}
            \Delta\omega & = &(\Delta \omega)_{\rm J}+(\Delta \omega)_{\rm Q},\label{eq:press1}\\
	    (\Delta\omega)_{\rm J} & = & \frac{4\pi a_\text{h}}{c^3q^{1/2}}\left[\frac{GM_{\rm h}\beta}{R_*(1+e)}\right]^{3/2},\label{eq:press2}\\
	    (\Delta\omega)_{\rm Q} & = & \frac{3\pi {a_\text{h}}^2}{c^4q^{2/3}}\left[\frac{GM_{\rm h}\beta}{R_*(1+e)}\right]^2\cos i,\label{eq:press3}\\
	    &&\nonumber\\
	    \Delta\Omega & = &(\Delta\Omega)_{\rm D} + (\Delta\Omega)_{\rm J} + (\Delta\Omega)_{\rm Q},\label{eq:press4}\\
	    (\Delta\Omega)_{\rm D} & = & \frac{6\pi GM_{\rm h}\beta}{c^2q^{1/3}R_*(1+e)},\label{eq:press5}\\
	    (\Delta\Omega)_{\rm J} & = & -4\cos i(\Delta\omega)_{\rm J}\label{eq:press6},\\
	    (\Delta\Omega)_{\rm Q} & = & \frac{1-5\cos^2i}{2\cos i}(\Delta\omega)_{\rm Q},\label{eq:press7}
	\end{eqnarray}
        where $\Delta\omega$ and $\Delta\Omega$ represent the orbit-averaged nodal and apsidal precessions. 
        The subscripts $\rm D$, $\rm J$, and $\rm Q$ correspond to the de Sitter, Lense-Thirring, and quadrupole terms, respectively. 
        The panels in Fig.~\ref{fig:precession} show $\Delta\omega$ and $\Delta\Omega$ as a function of the mass of the black hole $M_\text{h}$ and its spin $a_\text{h}$ for a penetration factor $\beta=1$ and inclination $i=45^{\circ}$. 
        Although not shown here the penetration factor can boost the effect of precession as most equations~\ref{eq:press1}-\ref{eq:press7} show a proportionality at least to first order in $\beta$. 
        However, deeper events are more likely to go directly inside the IBSO radius without having the chance to generate an electromagnetic signature (see Figure~\ref{fig:ibso}). 
        Thus, in these cases it is necessary to adopt a more careful approach to calculate the exact trajectories of the stream in the immediate vicinity of the black hole.

        The calculation continues until the first point in time at which the stream collides with itself. 
        Here, we consider that the event is triggered and material falls onto the black hole, forms a disc, and generates radiation and a disc wind (see Section~\ref{sec:tdemodel}). 
        In this way, by choosing the initial orbital parameters of the disrupted star and the parameters of the event, it is possible to describe the three-dimensional geometric structure of the tidal stream that wraps around the black hole. 
        Although this approach has limitations in capturing the precise trajectory of the stream in the vicinity of the black hole, we argue that it is reasonable to use it for initialising the density structure of the leftover of the stream. 
        The main reason for this lies in that most of the mass in the stream will be concentrated further away from the black hole, since this is the region with the longest free-fall timescale. 
        For a more detailed description of this problem, we refer the reader to the work by \cite{batra2023} that developed a detailed analysis on the of the evolution of the stream thickness and its trajectory by solving an approximate tidal equation.
    \subsection{TDE model}
    \label{sec:tdemodel}
        The model is based on a stellar tidal disruption assuming the inefficient accretion scenario following the description by \cite{metzger2016}. 
        It is important to remark that this approach is not affected by the so-called ``missing energy" problem, i.e. the small total energy ($\sim1~\text{per}~\text{cent}$) inferred from optical/UV observations relative to the expected energy released by the eventual accretion of half of a star. 
        An inefficient accretion model solves this problem by construction, since the total radiated energy $E_\text{tot}$ is small $E_\text{tot}<<f_\text{in}\eta(m_*/2)c^2$, where $f_\text{in}$ and $\eta$ the accretion fraction and the radiative efficiency, respectively. 
        Other solutions to this problem are: the absorption of most of the radiation, the fact that most of the energy could be in a jet beamed away from the line of sight to the Earth, or eccentric accretion discs that allow mass to fall directly to the black hole without circularisation and viscous accretion \citep[see][]{piran2015,svirski2017,lu2018}. 

        To introduce our mode, let us consider a star with mass $m_*=1~\text{M}_{\odot}$ and radius $R_*=1~\text{R}_{\odot}$, moving on a highly eccentric orbit $e=0.99$ around a black hole of mass $M_\text{h}$ and spin parameter $a_\text{h}$. 
        We fixed the penetration ratio to $\beta=1$, so that the orbital pericentre coincides with the tidal radius given by equation~\ref{eq:radius}. 
        Once the star is disrupted, the stellar stream wraps around the black hole due to its orbital precession until it collides with itself following the description given in Section~\ref{sec:analytical}. 
        In this model, we remain agnostic of the detailed evolution after this point until the electromagnetic signal and mass outflow are produced. 

        The accretion luminosity will be given by
	\begin{equation}
            \label{eq:luminosity}
	    L_\text{acc}(t) = \eta f_\text{in}f_\text{m}\dot{M}_\text{fb}(t)c^2,
	\end{equation}
	and a disc outflow with mass rate
	\begin{equation}
            \dot{M}_\text{w}(t)=(1-f_{\rm in})f_\text{m}\dot{M}_\text{fb}(t), 
	\end{equation}
        \noindent
        where $\dot{M}_\text{fb}(t)$ is the rate at which mass returns to pericentre after the disruption, $c$ is the speed of light, and $f_\text{m}$ is a free parameter of the model that indicates the fraction of the bound mass that is either accreted or launched in the wind, so that the rest ($1-f_\text{m}$) stays in the precessed stellar stream. 
        For instance, if $f_\text{m}=1$ no leftover material remains in the streams, so all the material is either accreted or ejected as a wind.
        Then, if only a small fraction of the material is accreted, i.e. $f_\text{in}\ll1$, the minimum speed for the material to become unbound at the tidal radius is given by
	\begin{equation}
            \label{eq:velmin}
            V_\text{w}^\text{min}=\sqrt{\frac{GM_\text{h}f_\text{in}}{(1-f_\text{in})r_\text{t}}}\approx9.8\times10^8\left(\frac{f_\text{in}}{0.1}\right)^{1/2}\left(\frac{M_\text{h}}{10^6~\text{M}_{\odot}}\right)^{1/3}~\text{cm\ s}^{-1},
	\end{equation}
        which corresponds to the initial velocity of the disc wind. 
        If the generated luminosity is capable of accelerating the wind, it could reach a maximum speed of
        \begin{equation}
            \label{eq:velmax}
            V_\text{w}^\text{max}\approx\sqrt{2\eta f_\text{in}}c=4.24\times10^9\left(\frac{\eta}{0.1}\right)^{1/2}\left(\frac{f_\text{in}}{0.1}\right)^{1/2}~\text{cm~s}^{-1}.
	\end{equation}
        Notice that for a more massive black hole the minimum wind speed will be closer to the maximum speed.
        Regardless of the black hole mass, relativistic effects should not be relevant since at most $V_\text{w}/c\approx0.14$, which implies a Lorentz factor of $\gamma\approx1.01$.

        We consider that both mass and luminosity are launched isotropically from the tidal radius $r_\text{t}$. 
        Although the sphericity of the wind and radiation of the event is an assumption, we argue that this is reasonable since it is possible for the accretion disc to precess on relatively short timescales \citep[e.g.][]{stone2012,franchini2016,teboul2023}.
        Additionally, we must assume that this process is fast enough for the luminosity and wind to encounter a significant fraction of mass of the stellar stream wrapped around the black hole. 
        We expect that the interaction of both the radiation and the wind with the surrounding structure will leave observable signatures on the light curves of these events, which will depend on the line of sight.      
\section{Numerical simulations}
\label{sec:numerical}
    \subsection{Equations}
    \label{sec:eqs}
        We perform two-dimensional $(r,\theta)$ radiation-hydrodynamic simulations using the moving-mesh hydrodynamic code JET \citep{duffell2011,duffell2013} with its radiation treatment module \citep{calderon2021}. 
        The code solves the radiation-hydrodynamic equations in the mixed-frame formulation following \cite{krumholz2007}. 
        The closure relation for the radiation treatment is the flux-limited diffusion approximation \citep{alme1973}. 
        Thus, the radiative flux is expressed as a function of the radiative energy gradient following Fick’s law. 
        Then, the set of equations in the mixed-frame formulation is
	\begin{eqnarray}
            \frac{\partial \rho}{\partial t} + \nabla\cdot\left(\rho\mathbf{u}\right) &=& 0,\\
		    \frac{\partial (\rho\mathbf{u})}{\partial t} + \nabla\cdot\left(\rho\mathbf{u}\mathbf{u}\right) + \nabla p &=&
		    -\lambda\nabla E_\text{r} ,\\
            \frac{\partial(\rho E)}{\partial t} + \nabla\cdot\left(\rho E\mathbf{u}+p\mathbf{u}\right)
            &=&
            -c\kappa_{\rm P}\left(a_{\rm r}T^4-E_\text{r}^{(0)}\right)-\lambda\mathbf{u}\cdot\nabla E_\text{r}\nonumber\\
            &&\\
		    \frac{\partial E_\text{r}}{\partial t} + \nabla\cdot\left(\frac{3-f}{2}E_\text{r}\mathbf{u}\right)
		    &=&
		    c\kappa_\text{P}\left(a_\text{r}T^4-E_\text{r}^{(0)}\right)+\lambda\mathbf{u}\cdot\nabla E_r \nonumber\\
		    &&
		    +\nabla\cdot\left(\frac{c\lambda}{\chi_\text{R}}\nabla E_\text{r}\right),
	\end{eqnarray} 
        \noindent where $\rho$, $\mathbf{u}$, $P$, and $E_\text{r}$ are the radiation-hydrodynamic variables of mass density, fluid velocity, thermal pressure, and radiation energy density, respectively. 
        The total specific matter energy density E is given by $\rho E=\rho\mathbf{u}\cdot\mathbf{u}/2+e$, where $e$ is the internal energy density. 
        The constants $c$ and $a_\text{r}$ are the light speed and the radiation constant, respectively. 
        The coefficients $\kappa_\text{P}$ and $\chi_\text{R}$ represent the Planck and Rosseland absorption coefficients (in units of inverse length), respectively. 
        The flux limiter $\lambda$ and the Eddington factor $f$ were set following \cite{levermore1981},  
        \begin{eqnarray}
            \lambda(R)&=&\frac{2+R}{6+3R+R^2},\\
            f(\lambda,R)&=&\lambda+\lambda^2R^2,
        \end{eqnarray}
        where
        \begin{equation}
            R=\frac{|\nabla E_r|}{\chi_{\rm R}E_r}.
        \end{equation}
        The flux limiter allows radiation to diffuse in the optically thick regime so that $\lambda\to 1/3$, while in the optically thin limit $\lambda\to 1/R$ so that radiation propagates with the speed of light. 
    	
        Quantities with the superscript $(0)$ are measured in the co-moving frame while the rest in the lab frame. 
        To relate $E_\text{r}$ in the co-moving to the lab frame we used the expression derived by \cite{zhang2011} under the flux-limited diffusion approximation
        \begin{equation}
            E_\text{r}^{(0)}=E_\text{r}+2\frac{\lambda}{\chi_\text{R}}\frac{\mathbf{u}}{c}\cdot\nabla E_\text{r}+\mathcal{O}(u^2/c^2).
        \end{equation}
        For further details of the implementation and method for solving the radiation hydrodynamic equations we defer the reader to the appendices A and B of \cite{calderon2021}.
        
        The simulations consider an adiabatic equation of state ($\gamma=5/3$), Solar composition ($X=0.7$, $Y=0.28$, $Z=0.02$), and a constant mean molecular weight assuming that the gas is fully ionised. 
        We set Planck and Rosseland opacities ($k_\text{R}$ and $k_\text{P}$) following \cite{metzger2017} \citep[see also][]{pejcha2017,matsumoto2022} by using the following analytical expressions
        \begin{eqnarray}
            k_\text{R}
            &=&
            k_\text{m}+[k_{\text{H}^-}+(k_\text{e}^{-1}+k_\text{K}^{-1})]^{-1},\\
            k_\text{P}
            &=&
            0.1k_\text{m}+(k_{\text{H}^-}+k_\text{K}^{-1})^{-1},
        \end{eqnarray}
        where
        \begin{eqnarray}
            k_\text{m}
            &=&
            0.1Z~\text{cm}^2~\text{g}^{-1},\\
            k_{\text{H}^-}
            &=&1.1\times10^{-25}Z^{0.5}\left(\frac{\rho}{\text{g}~\text{cm}^{-3}}\right)^{0.5}\left(\frac{T}{\text{K}}\right)^{7.7}~\text{cm}^{2}~\text{g}^{-1},\\
            k_\text{K}
            &=&4.0\times10^{25}(1.0+X)Z\left(\frac{\rho}{\text{g}~\text{cm}^{-3}}\right)\left(\frac{T}{K}\right)^{3.5}~\text{cm}^2~\text{g}^{-1},\nonumber\\
            &&\\
            k_\text{e}
            &=& 0.2(1.0+X)~\text{cm}^2~\text{g}^{-1}.
        \end{eqnarray}
        Here, $k_\text{m}$ $k_{\text{H}^-}$, $k_\text{e}$, and $k_\text{K}$ correspond to the contributions of molecular, $\text{H}^-$, electron scattering, and Kramer opacities, respectively.
        \begin{figure*}
            \centering
	    \includegraphics[width=0.95\linewidth]{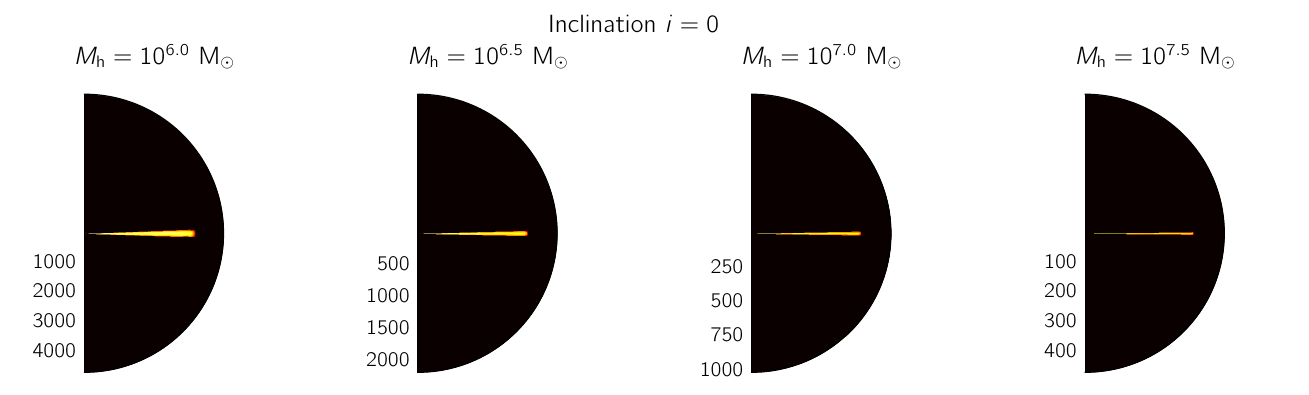}
	    \includegraphics[width=0.95\linewidth]{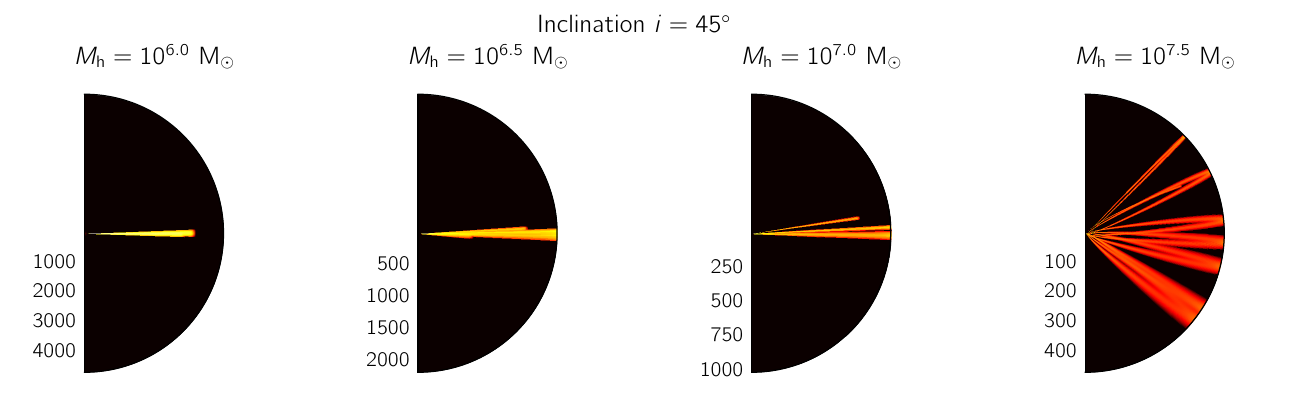}
            \includegraphics[width=0.95\linewidth]{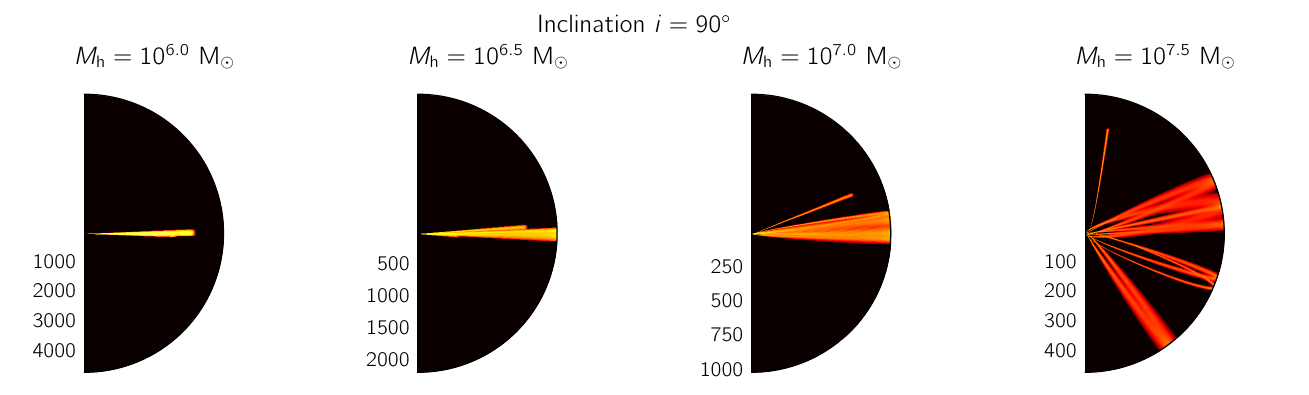}
            \includegraphics[width=0.5\linewidth]{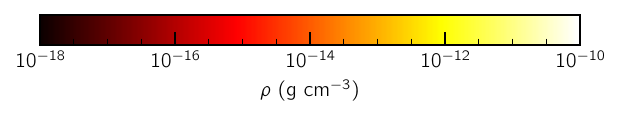}
            \caption{
            Two-dimensional $(r,\theta)$ density maps of the precessing TDE model with $a_\text{h}=0.9$ for varying inclination $i$ and black hole mass $M_\text{h}$. 
            Upper, central, and lower rows show models with inclination $i=0$, $45^{\circ}$, $90^{\circ}$, respectively. 
            Columns contain panels with a given black hole mass, from left- to right-hand side $M_\text{h}=10^{6.0}$, $10^{6.5}$, $10^{7.0}$, $10^{7.5}~\text{M}_{\odot}$. 
            Radial spatial scales are shown in units of Schwarzschild radii according to their black hole mass, i.e $R_\text{Sch}=2GM_{\rm h}/c^2$. 
            Notice that the larger the black hole mass the effect of relativistic precession is more relevant, hence the stellar stream wraps covering a larger solid angle.}
            \label{fig:ics}
        \end{figure*}
    \subsection{Initial conditions}
    \label{sec:ics}
        We set a two-dimensional spherical $(r,\theta)$ domain spanning radial and polar extensions of $(r_\text{t},100r_\text{t})$ and $(0,180^{\circ})$, respectively. 
        It is important to remark that the radial extension only refers to the initial state of the domain, as the moving-mesh aspect of the code will follow the expansion of the fluid.
        The density distribution is initialised to the precessed tidal stream following the analytical model presented in Section~\ref{sec:tdemodel}. 
        This is determined by the properties of the black hole: $M_\text{h}$, and $a_\text{h}$; of the star:  $m_*$ and $R_*$; and of the stellar orbit: $\beta$, $e$, and $i$. 
        Once we have calculated the structure of the stellar stream as a set of ellipses until an intersection is found, we estimate the density within the stream assuming that the amount of mass at a given location is proportional to the free-fall timescale at such a location. 
        We assumed that the stream has a cylindrical shape, whose radius is calculated using equation~\ref{eq:radius}. 
        Then, we normalise the density distribution so that the total mass in the stream is equivalent to $0.5(1-f_\text{m})m_*$.
        Then, we integrate it in the azimuthal direction in order to be able to represent it in the two-dimensional domain. 
        However, the complex geometry of the problem and the lack of symmetry complicates the choice of the direction of the symmetry axis. 
        Based on our tests, we have chosen the axis depending on every model, so that the density structure is located as close as possible to the direction $\theta=90^{\circ}$. 
        By doing so, we avoid the presence of numerical artifacts close to the symmetry axis as the density structure is as far as possible from it. 
        In particular, this choice allows us to simulate both the low and high inclination cases, since other choices, e.g. the axis aligned with the initial orbital angular momentum or the black hole spin direction, part of the structure will inevitably be at the poles.
        
        The stream is considered to be at rest with a low thermal pressure (temperature of $T=10^3~\text{K}$) and low radiation energy density ($E_\text{r}=10^{-16}~\text{erg}~\text{cm}^{-3}$). 
        The empty regions of the domain, i.e. where the stream did not pass, are filled with material with very low density, thermal pressure, and radiation energy density, so that they have negligible effect on the wind and luminosity injected as boundary conditions (see Section~\ref{sec:bcs}). 
        Fig.~\ref{fig:ics} presents four examples of the initial density distribution for different parameters. 
        Notice that the stream is confined to the orbital plane if the black hole mass is small, the stellar orbit and black hole spin are roughly aligned (see equations~\ref{eq:press1}-\ref{eq:press7}). 
        This is also the case if the black hole spin is smaller unless both inclination and black hole mass are large (see Appendix~\ref{sec:app}).
        On the other hand, if the inclination is nonzero and the black hole is more massive the precession takes the stream out of the orbital plane, which causes its spread over a larger solid angle (see midle and bottom right-hand side panels of Fig.~\ref{fig:ics}).
        \begin{figure}
            \centering
            \includegraphics[width=\linewidth]{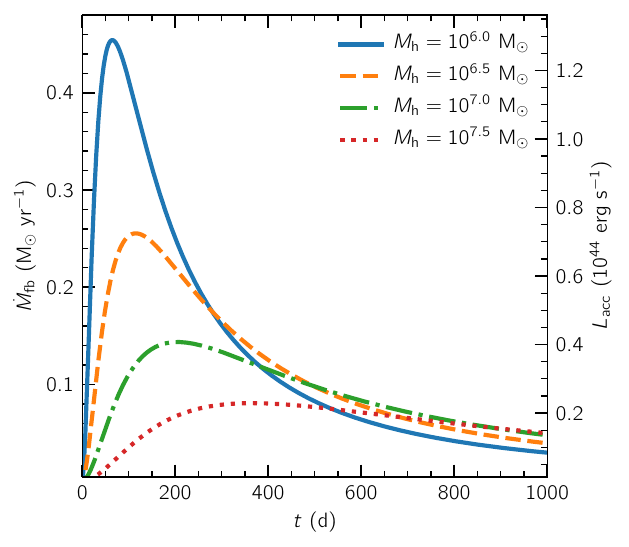}
            \caption{
            Fallback rates and accretion luminosity of a Solar-like star disrupted by a black hole of mass $10^6~\text{M}_{\odot}$ (solid blue line), $10^{6.5}~\text{M}_{\odot}$ (dashed orange line), $10^{7.0}~\text{M}_{\odot}$  (dash-dotted green line), and $10^{7.5}~\text{M}_{\odot}$ (dotted red line) with parameters $\beta=1$ and $e=0.99$. 
            The stellar density profile was modelled as a polytrope with $n=1.5$. 
            The fallback rate was computed using the approach by \protect\cite{lodato2009}.
	    The luminosity was computed using equation~\ref{eq:luminosity}.}
	    \label{fig:fallback}
	\end{figure}
        \begin{table}
            \centering
	    \begin{threeparttable}
	        \caption{Simulated models and parameters.}
	        \begin{tabular}{lccccc}
                  \hline
	             Model
	             & $\log\left(\frac{M_\text{h}}{M_{\odot}}\right)$
	             & $i$
	             & $\eta$
	             & $f_\text{m}$
	             & Init. Res.
	             \\
	             (1)&(2)&(3)&(4)&(5)&(6)\\
	             \hline
	             \hline
	             &&&&&\\
                  M60 & $6.0$ & $90^{\circ}$ & $0.1$ & $0.5$ & $512^2$\\
	             M65  & $6.5$ & $90^{\circ}$ & $0.1$ & $0.5$ & $512^2$\\
	             M70  & $7.0$ & $90^{\circ}$ & $0.1$ & $0.5$ & $512^2$\\
	             M75  & $7.5$ & $90^{\circ}$ & $0.1$ & $0.5$ & $512^2$\\
	             &&&&&\\ 
                  M60-$\eta$01 & $6.0$ & $90^{\circ}$ & $0.01$ & $0.5$ & $512^2$\\
	             M70-$\eta$01  & $7.0$ & $90^{\circ}$ & $0.01$ & $0.5$ & $512^2$\\
                 &&&&&\\  
                  M60-$i45$ & $6.0$ & $45^{\circ}$ & $0.1$ & $0.5$ & $512^2$\\
	             M70-$i45$  & $7.0$ & $45^{\circ}$ & $0.1$ & $0.5$ & $512^2$\\
                 &&&&&\\
                 M60-$i0$  & $6.0$ & $0$ & $0.1$ & $0.5$ & $512^2$\\
                 M70-$i0$  & $7.0$ & $0$ & $0.1$ & $0.5$ & $512^2$\\
                 &&&&&\\
                 M60-$f9$  & $6.0$ & $90^{\circ}$ & $0.1$ & $0.9$ & $512^2$\\
                 M70-$f9$  & $7.0$ & $90^{\circ}$ & $0.1$ & $0.9$ & $512^2$\\
                \hline
	        \end{tabular}
	        \label{tab:models}
	        \begin{tablenotes}
	            \item
	            \textbf{Notes.}
	            Column 1: the name of the model. 
	            Column 2: black hole mass $M_\text{h}$. 
	            Column 3: inclination $i$ between the orbital plane of the disrupted star and the black hole spin. 
	            Column 4: accretion efficiency $\eta$.  
	            Column 5: fraction of the bound mass that is either accreted or goes into the disc outflow $f_\text{m}$.  
	            Column 6: initial resolution in the radial $N_r$ and polar $N_{\theta}$ directions.
	        \end{tablenotes}
	    \end{threeparttable}
	    \end{table}
    \subsection{Boundary conditions and source terms}
    \label{sec:bcs}
        To include the wind and luminosity of the event in our simulation we set boundary conditions as source terms in the innermost radius of the domain. 
        The wind is injected as source terms in the matter, momentum, and energy equations, which is assumed to be time dependant and spherically symmetric. 
        The wind is specified by a mass-loss rate $\dot{M}_\text{w}$, $V_\text{w}$, $T_\text{w}$, and $E_\text{r,w}$. 
        The mass-loss rate of the wind is set to the fallback rate of a TDE, assuming that the total mass of the outflow corresponds to half of the stellar mass times $f_\text{m}$.
        The fallback rate is calculated using the method considering that the stellar structure is a polytrope before the disruption \citep{lodato2009}. 
        The resulting non-analytical mass-loss rate as a function of time was tabulated and used as a time dependant boundary condition.
        The velocity of the wind was set using equation~\ref{eq:velmin}, as this corresponds to the minimum value for launching an outflow from the tidal radius. 
        In principle, the luminosity may accelerate the outflow further but this will be taken into account self-consistently in the simulation.
        The temperature and radiation energy density of the wind were set to small enough values: $T_\text{w}=10^3~\text{K}$, and $E_\text{r,w}=10^{-16}\text{erg}~\text{cm}^{-3}$, so that the total energy of the wind is dominated by its kinetic energy.
        
        The luminosity was included as a boundary condition in the implicit step of the solver through the innermost boundary of the domain. 
        This was estimated using the fallback rate through equation~\ref{eq:luminosity}, and therefore is a time-dependent function. 
        As our code allows the motion of the domain along the radial direction we had to take into account potential changes in the radius where both the wind and irradiation are injected. 
        This was handled considering that the wind is continuously blown from a fixed radius, and tracking the potential time difference it takes to arrive to the innermost radial boundary. 
        We also ensured that the stream structure was far enough to this boundary throughout the whole simulation. 
        Additionally, we took into account energy degradation due to adiabatic expansion only if the wind was dense enough for trapping the radiation in it. 
        This was taken into account adding an extra factor $(r_\text{tr}/r_\text{in})^{-2/3}$ to the injected luminosity, where $r_\text{tr}$ corresponds to the trapping radius, i.e. the radius at which the diffusion timescale equates the dynamical timescale \citep{piro2020,calderon2021}. 
    \subsection{Models}
    \label{sec:models}
        We investigate models whose parameters would maximise the deviations from the standard scenario.  
        Table~\ref{tab:models} shows the parameters of the 12 models explored in this work. 
        All models consider a Solar-like star, i.e. $m_*=1~\rm M_{\odot}$ and $R_*=1~\rm R_{\odot}$, being disrupted by a black hole with a spin parameter $a_\text{h}=0.9$ in an encounter with $\beta=1$, and assuming that a small fraction of the bound material is accreted $f_\text{in}=0.1$. 
        The choice of fixing the spin parameter to high value was motivated due to the fact that smaller values cause only small precession out of the orbital plane. 
        As a result, the stream remains roughly aligned with it.  
        This fact is discussed further in the Appendix~\ref{sec:app}. 
        Thus, the parameters studied are black hole mass $M_\text{h}$, inclination of the orbital and black hole spin inclination $i$, accretion efficiency $\eta$, and stream mass fraction $(1-f_\text{m})$. 
        It is important to remark that we ensured that the pericentre of the disrupted star was larger than the corresponding IBSO radius in every model.
        The resolution of the simulations initially considered 512 spatial elements in each dimension logarithmically and linearly spaced along the radial and polar directions, respectively. 
        However, as the moving mesh of JET adapts during the simulation the typical resolution along the radial direction was about $\sim$1000 elements.
        Each model was simulated for at least 1,000 days, which had a computational cost of approximately 9,000-18,000 cpu~hours depending on the parameters of the model. 
        The calculations were performed at the Barbora cluster at the IT4Innovations National supercomputer center in Ostrava, Czech Republic\footnote{\url{https://www.it4i.cz/en/infrastructure/barbora}}.
        \begin{figure*}
        \begin{subfigure}[h]{.475\textwidth}
            \includegraphics[width=\linewidth]{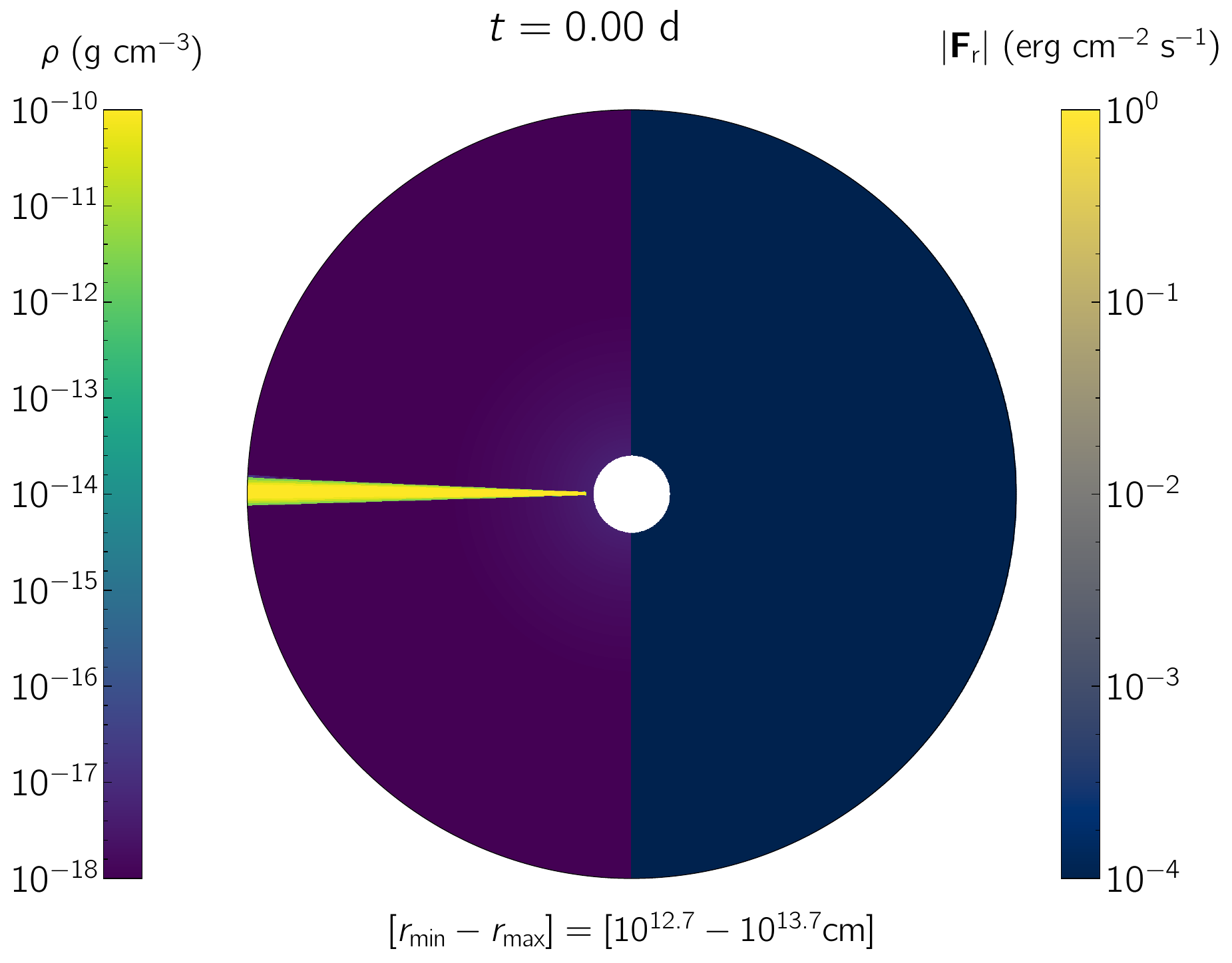}
            \caption{}
            \label{fig:m60_maps_a}
        \end{subfigure}
        \begin{subfigure}[h]{.475\textwidth}
            \includegraphics[width=\linewidth]{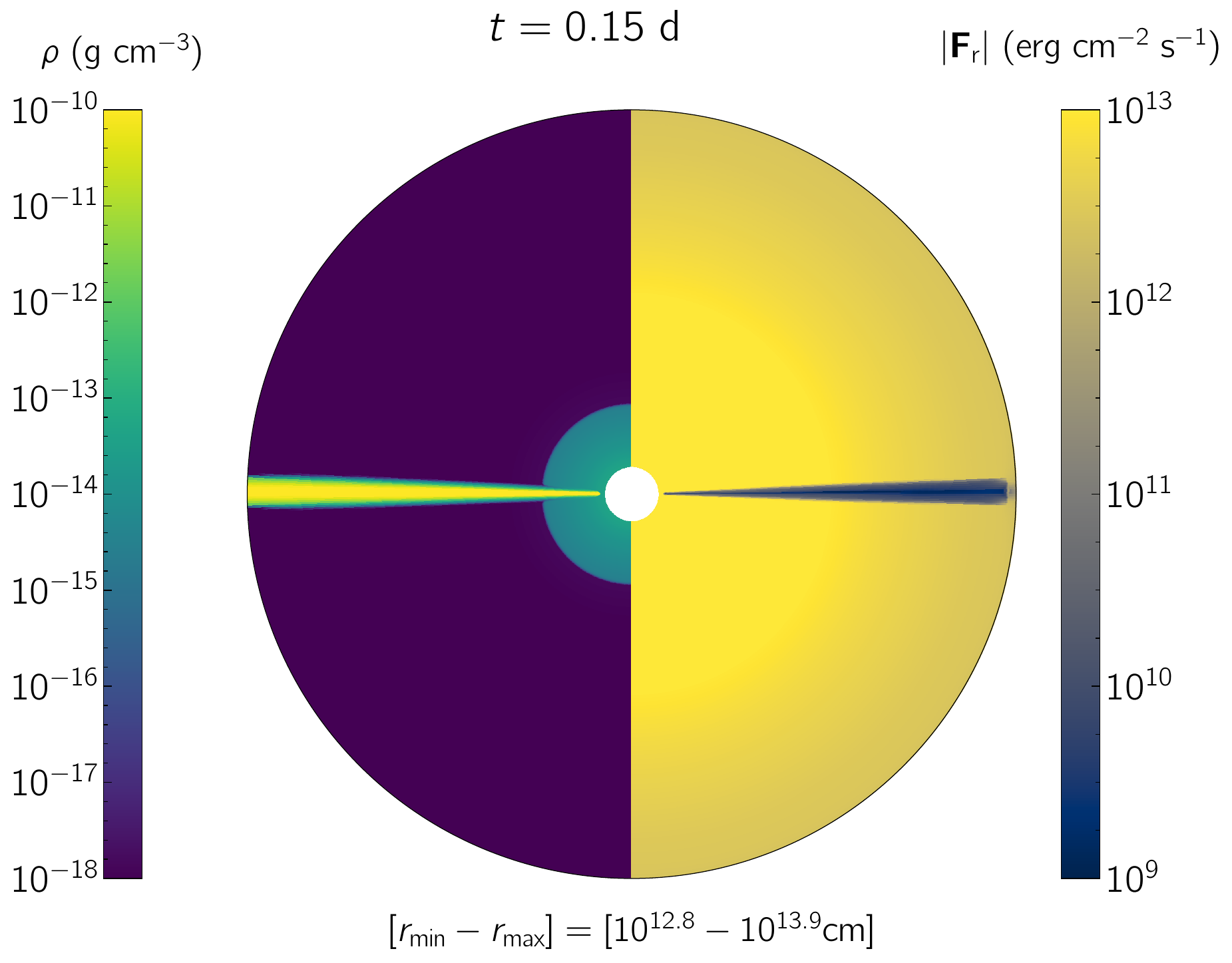}
            \caption{}
            \label{fig:m60_maps_b}
        \end{subfigure}
        \begin{subfigure}[h]{.475\textwidth}
            \includegraphics[width=\linewidth]{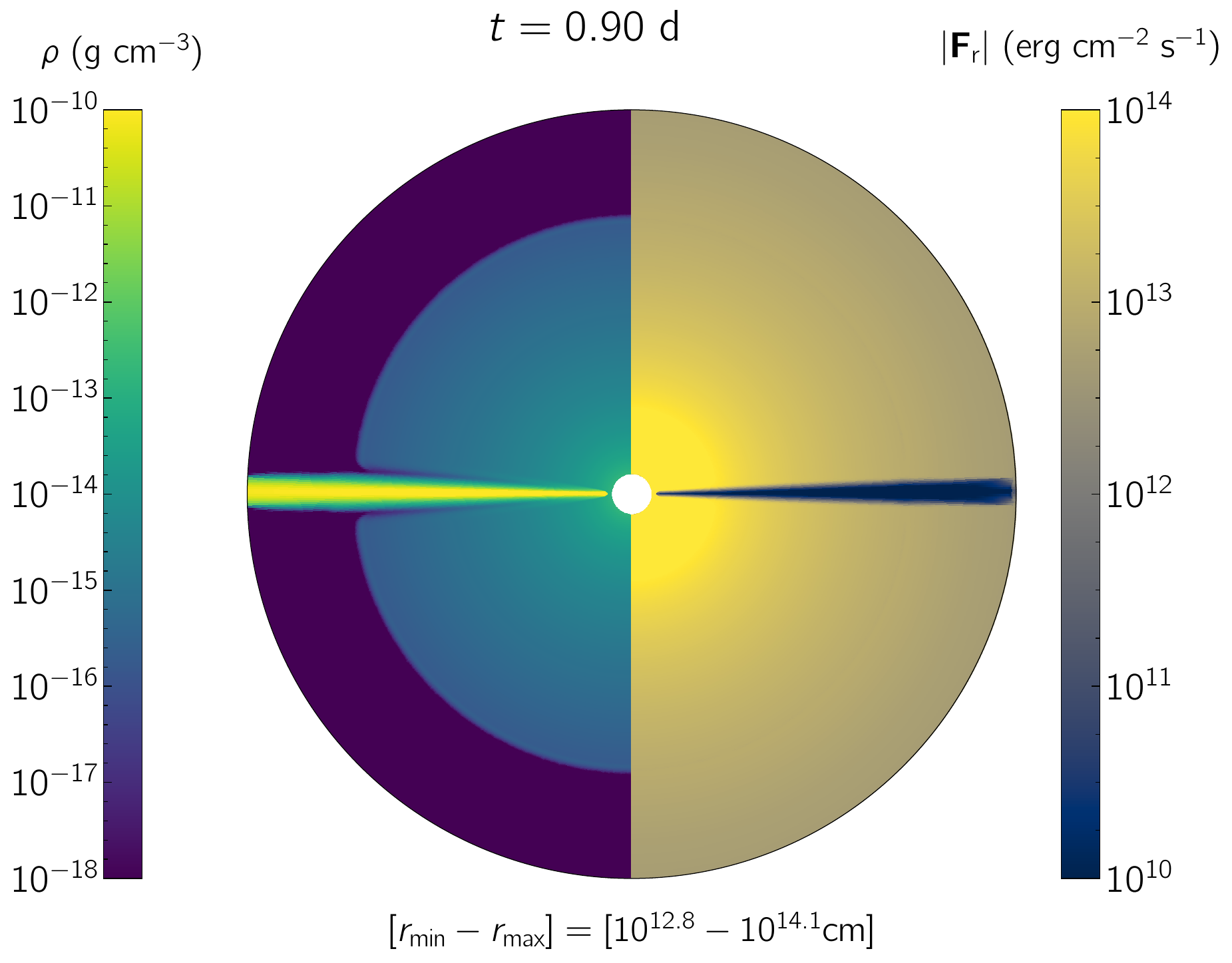}
            \caption{}
            \label{fig:m60_maps_c}
        \end{subfigure}
        \begin{subfigure}[h]{.475\textwidth}
            \includegraphics[width=\linewidth]{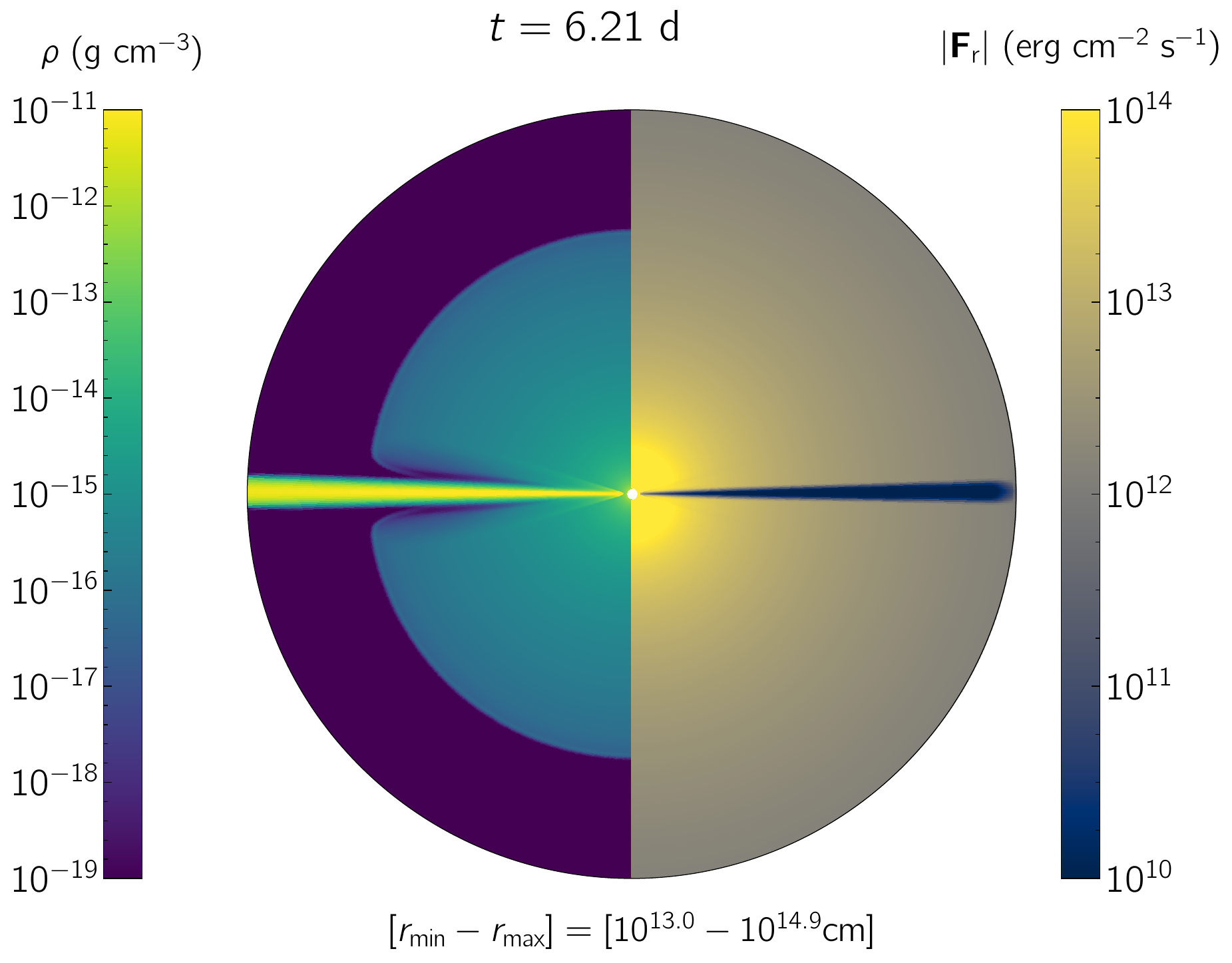}
            \caption{}
            \label{fig:m60_maps_d}
        \end{subfigure}   
        \begin{subfigure}[h]{.475\textwidth}
            \includegraphics[width=\linewidth]{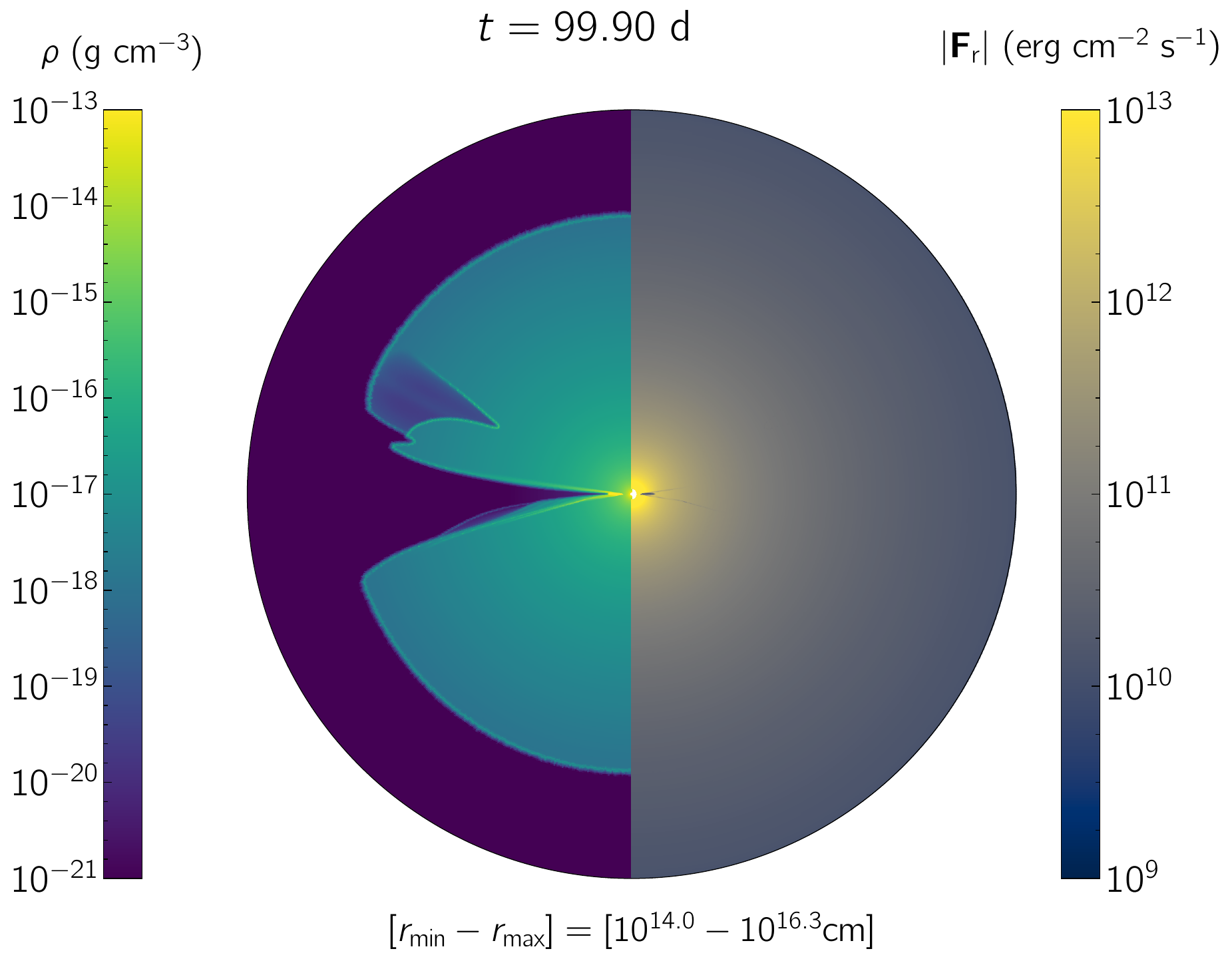}
            \caption{}
            \label{fig:m60_maps_e}
        \end{subfigure}
        \begin{subfigure}[h]{.475\textwidth}
            \includegraphics[width=\linewidth]{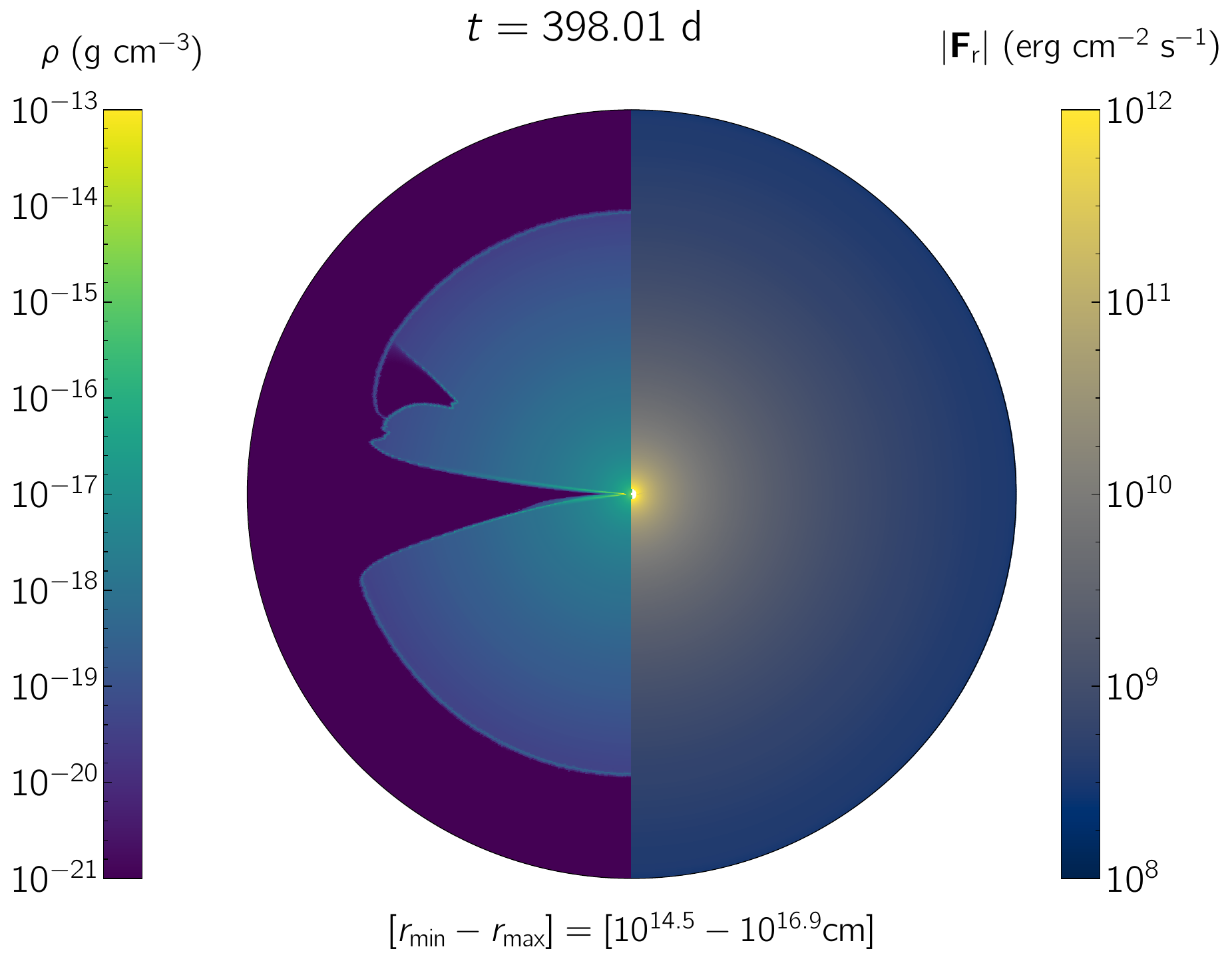}
            \caption{}
            \label{fig:m60_maps_f}
        \end{subfigure}   
        \caption{
        Evolution of density and radiative flux maps of model M60. 
        Each panel contains the entire two-dimensional domain duplicated in order to show density $\rho$ and magnitude of radiative flux $|\mathbf{F}_\text{r}|$ on a single panel. 
        The minimum and maximum radial extensions of the domain at a given time, i.e. $r_\text{min}$ and $r_\text{max}$ are included below each panel.
        From the upper left to the lower right panel, the simulation time of each panel are $t=0, 0.15, 0.9, 6.21, 99.9, 398.01~\text{d}$. 
        The asymmetric feature observed above the equator is a result of the precession that changes the density distribution of the stream, since the model M60-$i0$ remains symmetric (see animations in Supplementary Material).
        }
        \label{fig:m60_maps}
    \end{figure*}
\section{Results}
\label{sec:results}
    In this section, we present the results of the simulations discussing their evolution, the light curves generated, and describing the impact of the main parameters studied. 
    The fiducial set of simulations consider $M_\text{h}=10^{6.0},~10^{6.5},~10^{7.0},~10^{7.5}~\text{M}_{\odot}$, $i=90^{\circ}$, $\eta=0.1$, and $f_\text{m}=0.5$. 
    Thus, we describe these models, and then we discuss the impact of modifying these parameters.
        \begin{figure*}
            \centering
            \includegraphics[width=\linewidth]{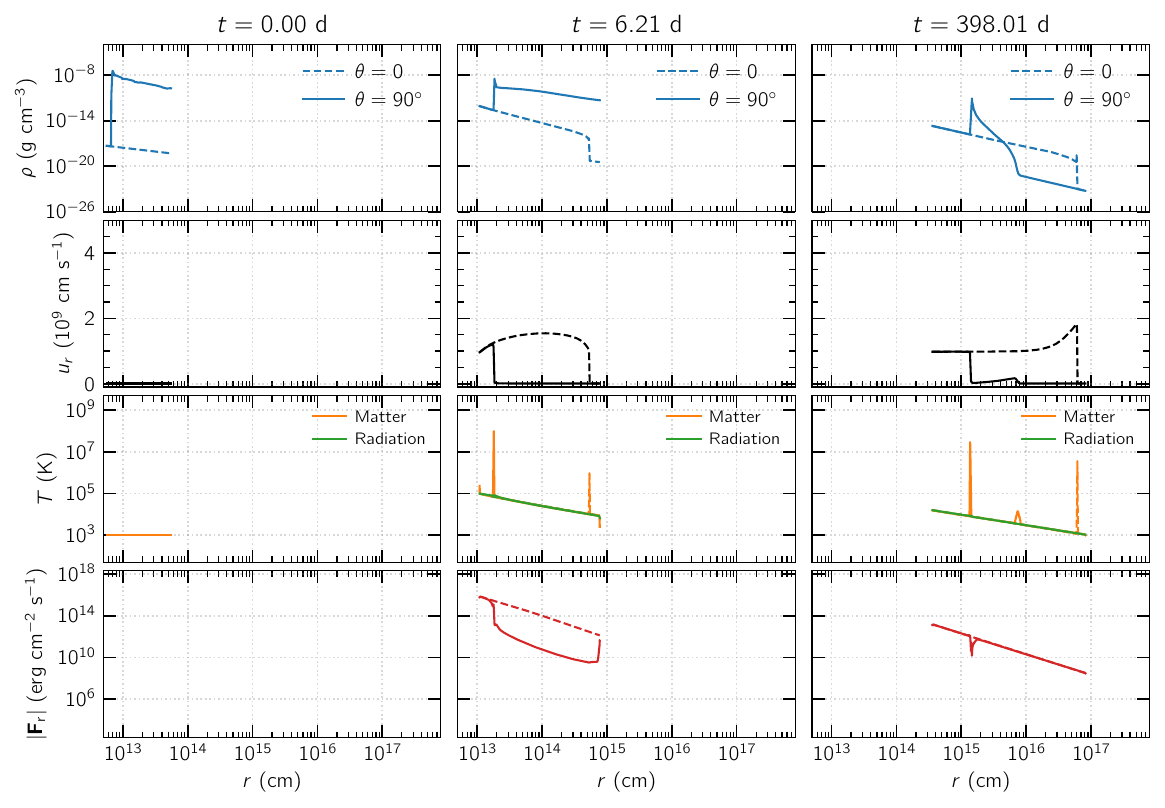}
            \caption{
            Radial profiles of density (upper row), radial velocity (second row), temperature (third row), and radiative flux (lower row) of model M60 along the polar and equatorial directions. 
            Dashed and solid lines show the variables for $\theta=0$ and $\theta=90^{\circ}$, respectively. 
            Panels along the same column show the variables at simulation times $t=0,6.21,398.01~\text{d}$, which coincide with the maps of Figs.~\ref{fig:m60_maps_a},~\ref{fig:m60_maps_d}, and~\ref{fig:m60_maps_f}, respectively. 
            Matter and radiation temperatures are displayed in orange and green lines, respectively.
            Notice how the domain samples different spatial regions throughout the simulation due to JET moving-mesh capability. 
            At $t=0$ the radiative flux is zero across the domain.
            }
            \label{fig:m60_ray_evol}
        \end{figure*}
    \subsection{Radiation hydrodynamic time evolution}
    \label{sec:tevol}
        \subsubsection{Model M60}
            The complete evolution of the model M60 is shown through density $\rho$ and module of radiative flux $|\mathbf{F}_\text{r}|$ maps in Fig.~\ref{fig:m60_maps}. 
            Each panel contains twice the complete domain of the simulation (left- and right-hand sides of the circles), in order to display both density and radiative flux simultaneously. 
            The quantities are shown in logarithmic scale while the spatial scale is linear with its minimum and maximum extensions, i.e. $r_\text{min}$ and $r_\text{max}$, which are shown in the lowermost region of each panel. 
            Figs.~\ref{fig:m60_maps_a}-\ref{fig:m60_maps_f} display maps at times $t=0,~0.15,~0.9,~6.21,~99.9,~398.01~\text{d}$. 
            Notice that both the spatial and color scales change significantly over time due to the radial motion of the mesh. 
        \begin{figure*}
        \begin{subfigure}[!t]{.475\textwidth}
            \includegraphics[width=\linewidth]{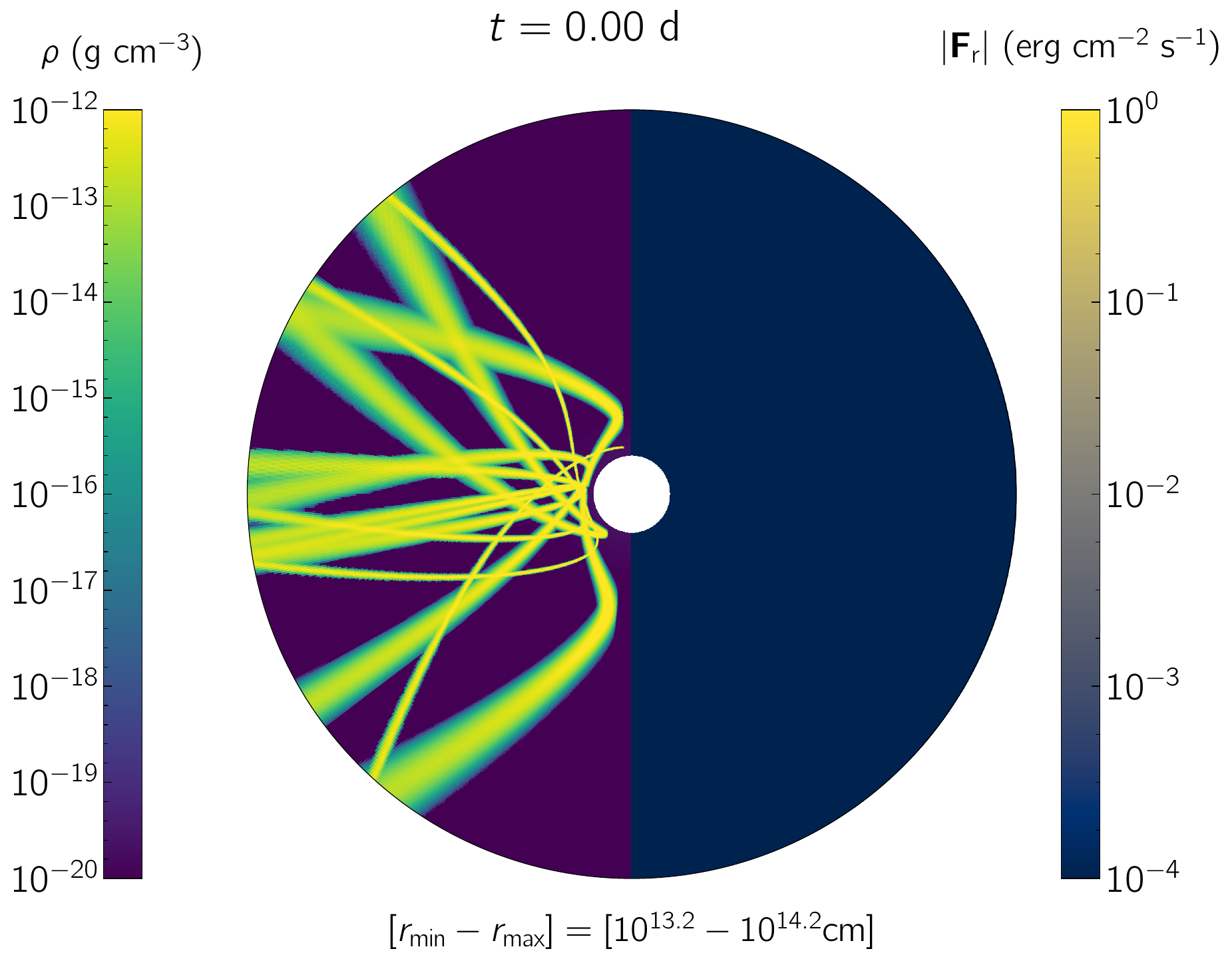}
            \caption{}
            \label{fig:m75_maps_a}
        \end{subfigure}
        \begin{subfigure}[!t]{.475\textwidth}
            \includegraphics[width=\linewidth]{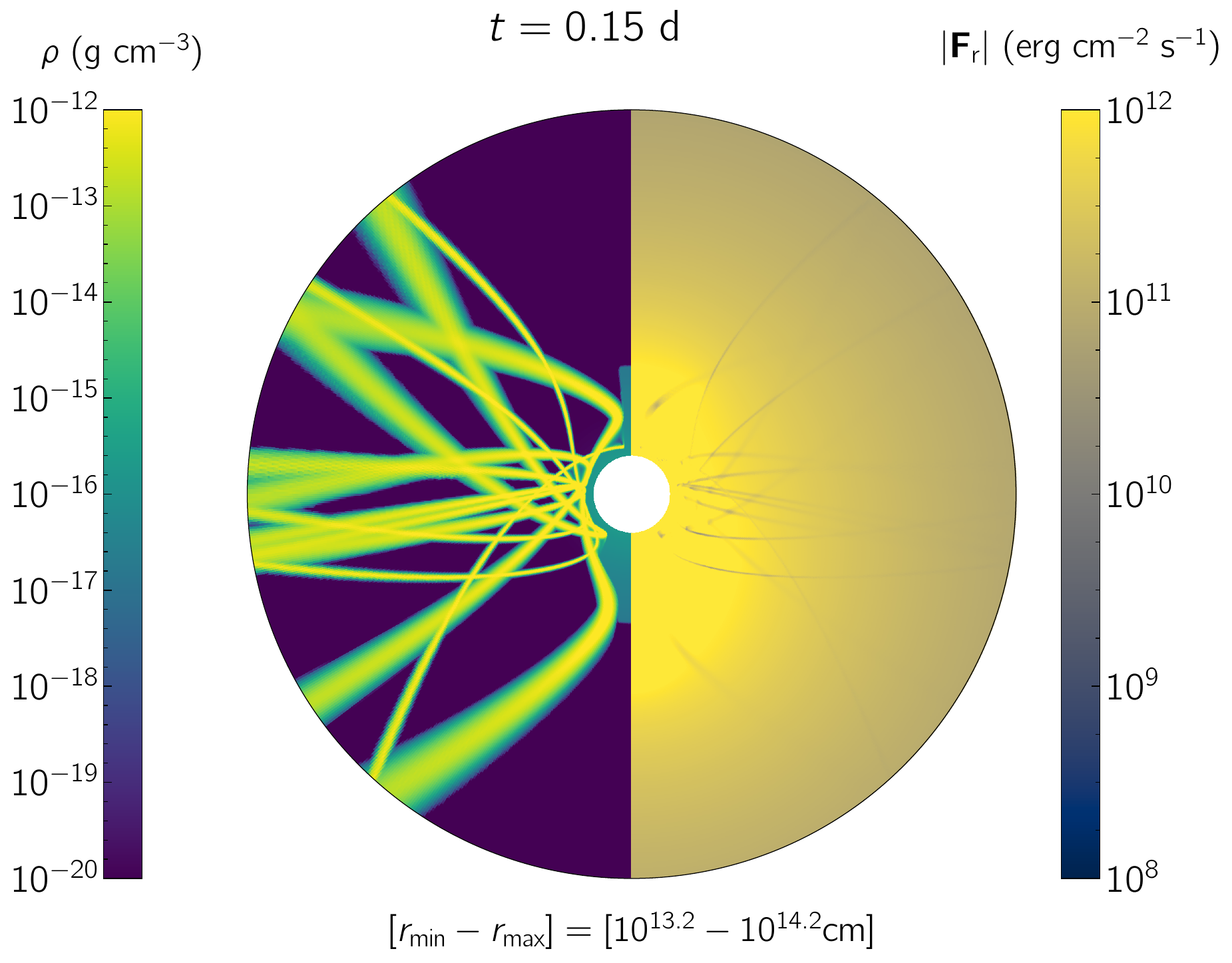}
            \caption{}
            \label{fig:m75_maps_b}
        \end{subfigure}
        \begin{subfigure}[!t]{.475\textwidth}
            \includegraphics[width=\linewidth]{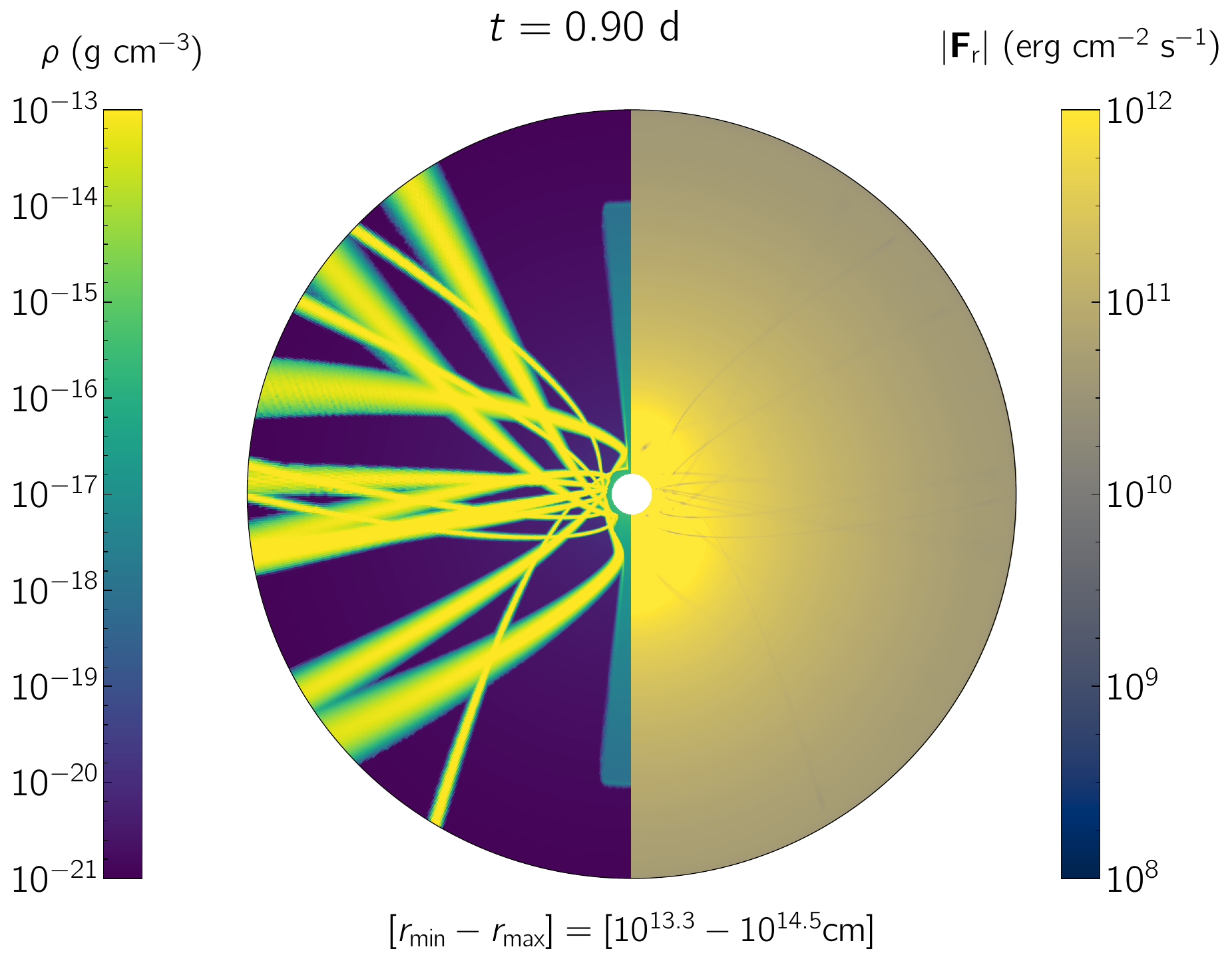}
            \caption{}
            \label{fig:m75_maps_c}
        \end{subfigure}
        \begin{subfigure}[!t]{.475\textwidth}
            \includegraphics[width=\linewidth]{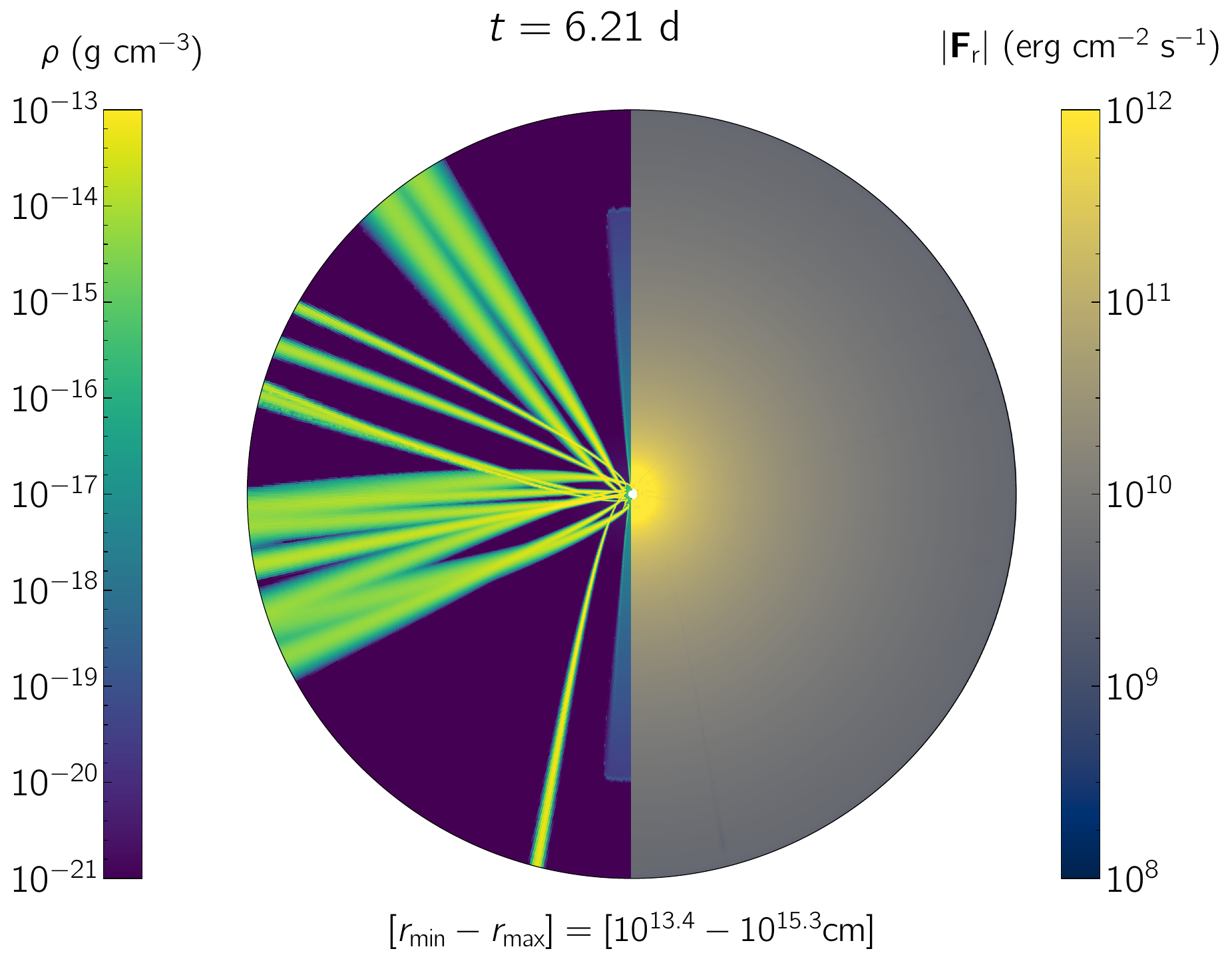}
            \caption{}
            \label{fig:m75_maps_d}
        \end{subfigure}   
        \begin{subfigure}[!t]{.475\textwidth}
            \includegraphics[width=\linewidth]{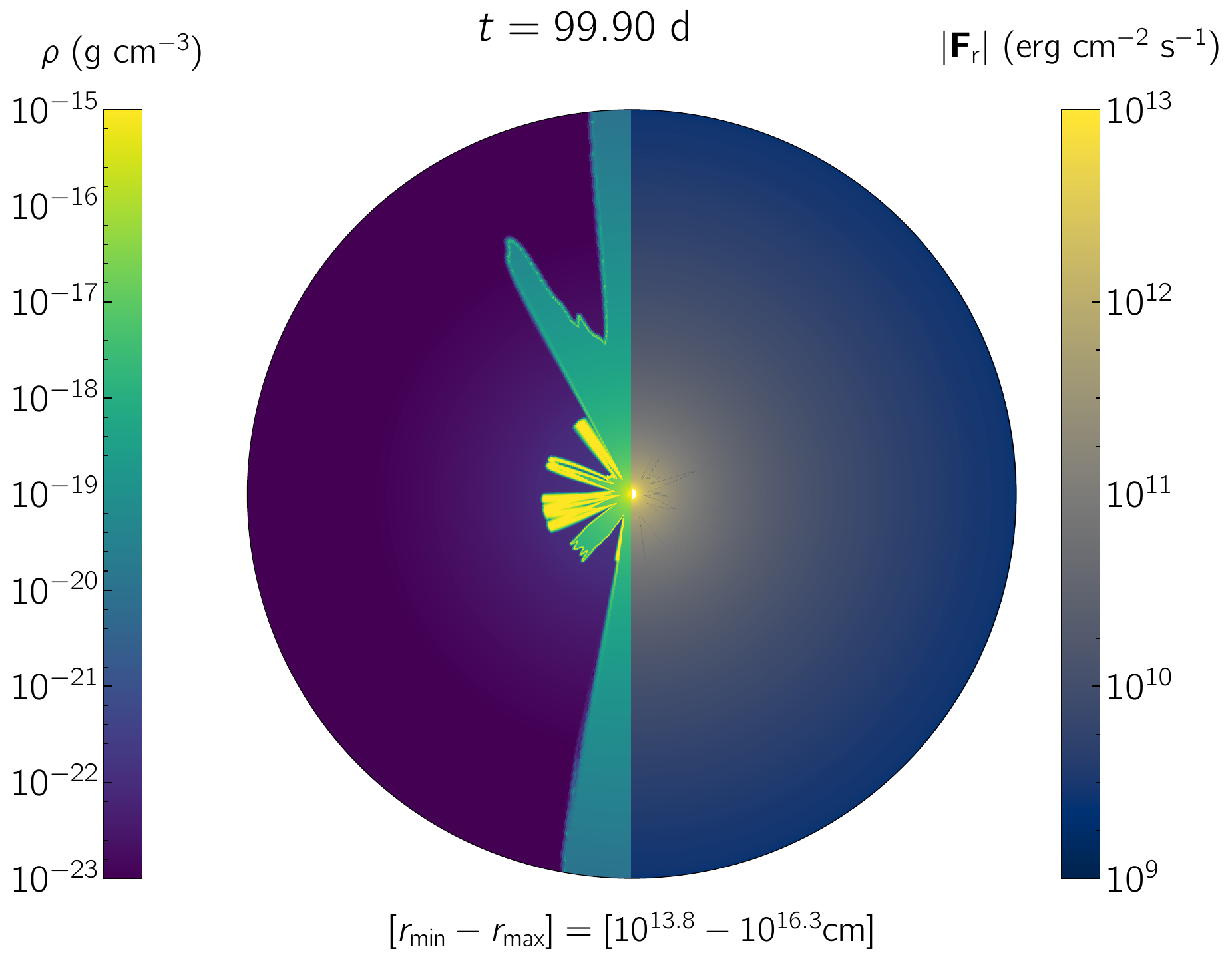}
            \caption{}
            \label{fig:m75_maps_e}
        \end{subfigure}
        \begin{subfigure}[!t]{.475\textwidth}
            \includegraphics[width=\linewidth]{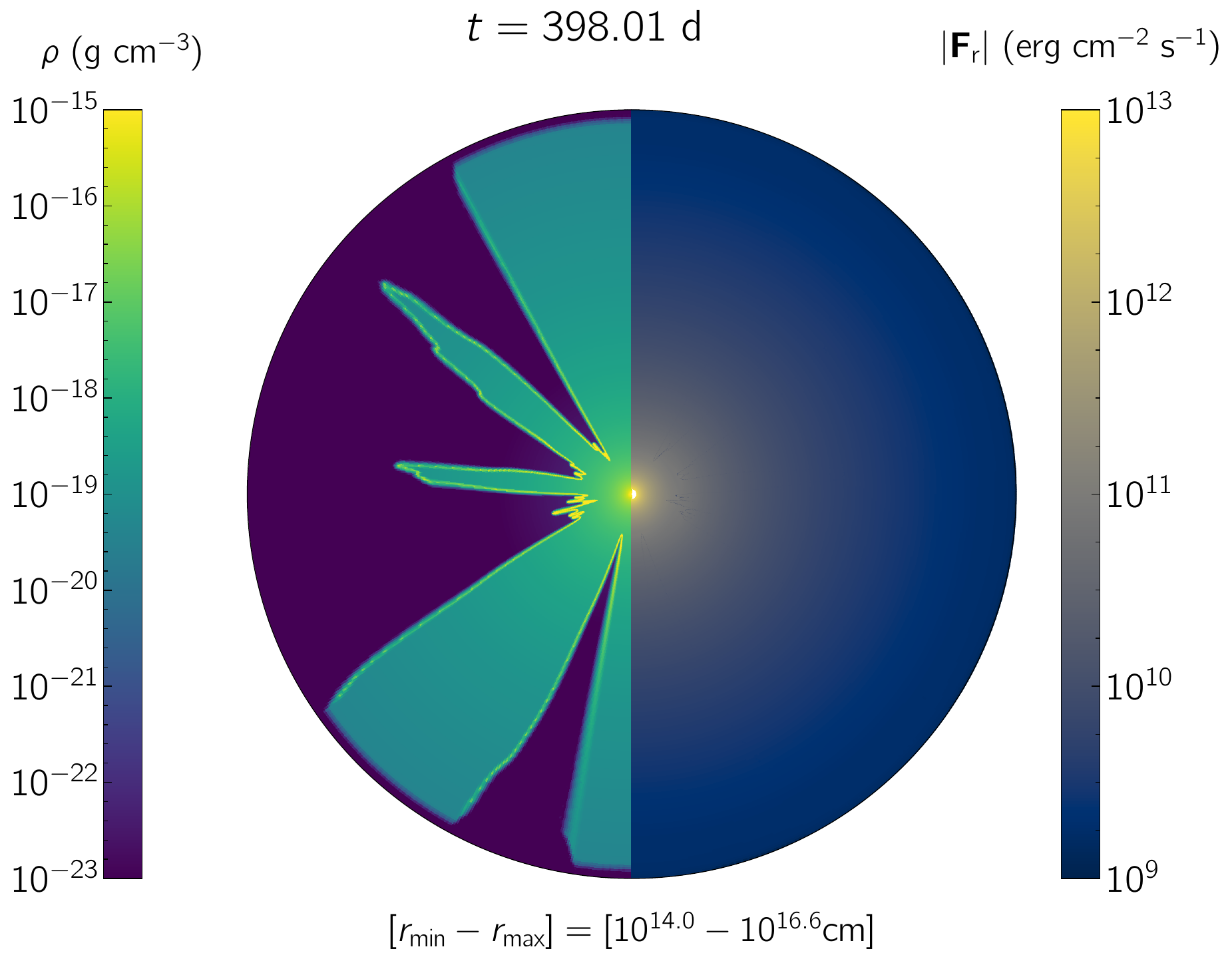}
            \caption{}
            \label{fig:m75_maps_f}
        \end{subfigure}   
            \caption{
            Analogous to Fig.~\ref{fig:m60_maps} but for model M75, i.e. with a black hole of $M_\text{h}=10^{7.5}~\text{M}_{\odot}$. 
            }
            \label{fig:m75_maps}
        \end{figure*}
        \begin{figure*}
            \centering
            \includegraphics[width=\linewidth]{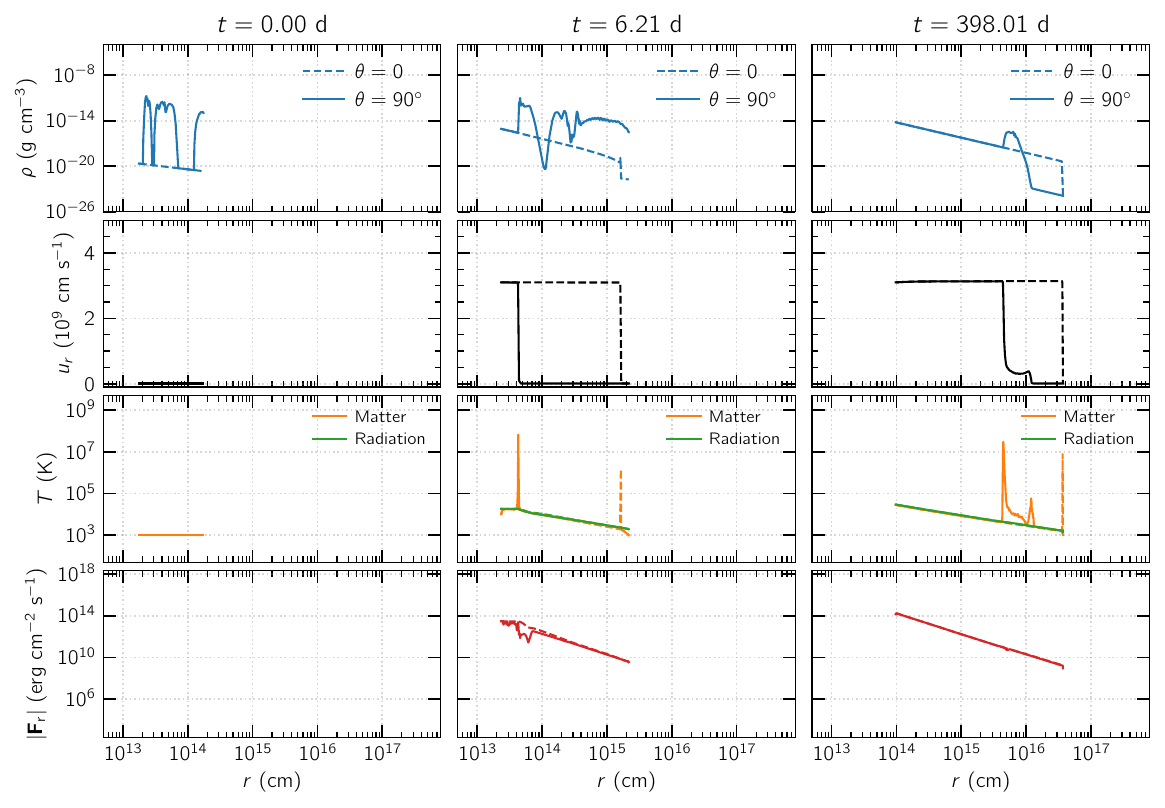}
            \caption{
            Analogous to Fig.~\ref{fig:m60_ray_evol} but for model M75, i.e. with a black hole of $M_\text{h}=10^{7.5}~\text{M}_{\odot}$.
            }
            \label{fig:m75_ray_evol}
        \end{figure*}
            At the beginning of the simulation ($t=0$), the density is negligible in the medium except along the orbital plane of the disrupted star, where the density is more than ten orders of magnitude larger (see Fig.~\ref{fig:m60_maps_a}). 
            As the radiation energy density was initially set to a low value the magnitude of the radiative flux is very small as well ($<10^{-4}~\text{erg}~\text{cm}^{-2}~\text{s}^{-1}$).
            Once the simulation starts, it can be seen how the wind is being launched from the inner boundary, although it is still less dense than the tidal stream (see Fig.~\ref{fig:m60_maps_a}). 
            The injected radiation manages to propagate quickly filling most of the domain, as at this stage the wind is completely optically thin. 
            However, this is not the case along the stellar stream, where radiation is forced to diffuse through it at a much slower speed, which can be seen as a shadow in the radiative flux map (see Fig.~\ref{fig:m60_maps_b}). 
            Over timescales of a couple to ten days, the evolution remains stationary (see Figs.~\ref{fig:m60_maps_c} and~\ref{fig:m60_maps_d}), i.e. the wind keeps being blown and travels outwards without impediment.  
            Despite its collision with the stream the ram pressure of the wind is not strong enough to accelerate it. 
            In this regime, the radiation has not had enough time to penetrate the dense stream yet and the absence of radiation in the stream is still noticeable. 
            At $t\approx100~\text{d}$, both the wind density and luminosity are close their maximum value given by the peak in the fallback rate (see Fig.~\ref{fig:fallback}). 
            As a result, the wind is much denser and has enough ram pressure to push the stream outwards, although not as fast as the free wind. 
            This can be seen as a bipolar outflow with a wide opening angle, deviating from a spherically symmetric wind (see Fig.~\ref{fig:m60_maps_e}). 
            Nevertheless, the radiative flux is isotropic at large scale as radiation has managed to diffuse isotropically regardless the presence of the stream, which at this scale is significantly smaller. 
            At longer timescales ($t\gtrsim400~\text{d}$), radiation has already filled the dense stream and the map looks completely isotropic (see Fig.~\ref{fig:m60_maps_f}). 
            In summary, the structure of the outflow is affected by the presence of the stream over the entire evolution but the radiation is only affected during the early phase of the event ($t<100~\text{d}$). 

            To analyse the evolution along different lines of sights we present radial profiles of the most relevant variables along $\theta=0$ (polar direction) and $\theta=90^{\circ}$ (equatorial direction) in Fig.~\ref{fig:m60_ray_evol}, which are shown as dashed and solid lines, respectively. 
            From upper- to bottom-most each row of panels show density, radial velocity, temperature, and magnitude of radiative flux, while from left- to right-hand side columns display same simulation time $t=0,~6.21,~398.01~\text{d}$. 
            Notice that the chosen sampling times coincide with the maps in Figs.~\ref{fig:m60_maps_a},~\ref{fig:m60_maps_d}, and~\ref{fig:m60_maps_f}. 
            These allow us to illustrate the phases of the evolution for observers along two extreme lines of sight. 
            Initially, the density profile is very different along both lines of sight due to the presence of the stream along $\theta=90^{\circ}$ (see uppermost left-hand side panel of Fig.~\ref{fig:m60_ray_evol}). 
            The rest of the variables were initialised identical along $\theta$: radial velocity and radiative flux are negligible in the initial conditions but both temperatures, matter and radiation, are in equilibrium at $10^3~\text{K}$ (see left-hand side column of Fig.~\ref{fig:m60_ray_evol}), where the temperature of radiation is defined as $T_\text{r}=(E_\text{r}/a_\text{r})^{1/4}$. 
            After a couple of days, differences can be spotted along the lines of sight (see central column of Fig.~\ref{fig:m60_ray_evol}). 
            Shocks develop at different locations depending on the line we consider. 
            Along $\theta=0$, the free expansion of the wind at high speed compresses the background material and heats it up to $10^6$-$10^7~\text{K}$. 
            The location of the shock is at $\sim5\times10^{14}~\text{cm}$ at $t=6.21~\text{d}$ and can be seen as a sudden drop in the density profile. 
            However, the encounter of the wind with the dense stream heats up the material in its inner boundary to $\sim10^8~\text{K}$ that cannot propagate as fast as the free wind region. 
            Additionally, it is possible to see that the outermost part of the wind along $\theta=0$ has been accelerated due to the action of the radiation at earlier times. 
            This also can be seen along $\theta=90^{\circ}$ but to less extent due to the presence of the stream.
            Only at later times ($t\gtrsim400~\text{d}$), the stream has been displaced outwards significantly (see right-hand column of Fig.~\ref{fig:m60_ray_evol}). 
            This has caused the stream to expand, decrease its density and its outermost parts have become optically thin. 
            Here, it can be observed that only in the densest region of the stream the radiative flux decreases but the rest coincides with the radiative flux along $\theta=0$. 
            Therefore, at this phase the source has become isotropic despite the presence of the remnant of the stellar stream. 
        \subsubsection{Model M75}
            The evolution of density and radiative flux of the model M75 is shown in Fig.~\ref{fig:m75_maps}, which is completely analogous to Fig.~\ref{fig:m60_maps}. 
            The main difference in this setup corresponds to the mass of the black hole that impacts directly the density distribution in the initial condition (see Fig.~\ref{fig:ics}) as well as the wind and radiation injected (see Fig.~\ref{fig:fallback}). 
            Fig.~\ref{fig:m75_maps_a} displays the initial state of the simulation, where the complex density structure is a product of the precession of the stellar stream. 
            Notice that the domain samples a larger spatial scale, as the massive black hole has a larger tidal radius. 

            Once the simulation starts, most of the wind interacts with the surrounding structure, as it covers a significant fraction of the solid angle (see Fig.~\ref{fig:m75_maps_b}). 
            The radiation fills the entire domain very quickly, even penetrating and going through the stellar stream. 
            This is a consequence of the material being more diluted at larger scales, as the total mass in the stream is the same but distributed over a bigger volume. 
            Therefore, more coverage of the solid angle due to the stream implies less dense regions in the stream. 
            Nevertheless, radiation still needs to diffuse at a slower speed when passing through the stream as its structure still can be recognized in the radiative flux maps, which implies an attenuation of the escaping radiation from the system (see Fig.~\ref{fig:m75_maps_b} and~\ref{fig:m75_maps_c}). 
            At longer times, the wind has expanded to large radii along the poles but the stream remains unaffected at larger scales due to the weak ram pressure of the wind (see Fig.~\ref{fig:m75_maps_d}). 
            In this case, the fallback rate is smaller and reaches its peak at later times ($t\approx300~\text{d}$). 
            Thus, significant hydrodynamic impact of the wind on the stream takes place at later times. 
            Indeed at $t\gtrsim100~\text{d}$, the wind manages to push the stream outwards to larger scales, yet the overall structure is closer to a bipolar configuration rather than to a spherical structure (see Fig.~\ref{fig:m75_maps_e}). 
            At this phase, the wind and stream are not dense enough to impact significantly the spherical symmetry of the radiative flux at these scales. 
            In the final phase ($t\gtrsim400~\text{d}$), the stream continues to travel outwards due to the action of the wind. 
            As a result, this is diluted even further and the source can be seen completely isotropic. 

            The detailed evolution of the model along extreme lines of sight is shown in Fig.~\ref{fig:m75_ray_evol}, which is also analogous to Fig.~\ref{fig:m60_ray_evol}. 
            The main features of the evolution can be seen in the density and radiative flux evolution, displayed in the upper- and bottom-most rows. 
            The density profiles along $\theta=0$ and $\theta=90^{\circ}$ show the former as a simple low density decay and the latter as a much denser structure caused by the precessed stream. 
            Once wind and radiation are injected through the innermost boundary, the wind tries to sweep the medium despite not having enough momentum for displacing the stream during the initial phase. 
            Regardless, shocks are formed at the maximum extension of the wind, which can be seen in the middle column of Fig.~\ref{fig:m75_ray_evol}: at $r\approx10^{15}~\text{cm}$ along $\theta=0$ and $r\approx5\times10^{13}~\text{cm}$ along $\theta=90^{\circ}$. 
            In the density profiles, this is seen as a small increase in the density, while in the temperature profiles it is seen as sharp increases in the matter temperature. 
            Notice how the velocity structure is different along each line of sight. 
            The free expansion of the wind along $\theta=0$ allows a constant speed of the wind but along $\theta=90^{\circ}$ the wind stopped moving outwards due to the inertia of the stream. 
            At this stage, the stream manages to attenuate the radiation, as it needs to diffuse in order to go through it. 
            However, this only occurs at the parts of the stream that are close to pericentre, i.e. at smaller scales as the stream is not as dense at further radii. 
            This can be observed as the radiative flux along $\theta=0$ and $\theta=90^{\circ}$ tends to agree for $r\gtrsim10^{14}~\text{cm}$ at this stage. 
            In the last phase ($t\gtrsim400~\text{d}$), the wind has swept the material in the stream along $\theta=90^{\circ}$ and formed a dense, thick shell kept at high temperature (see right-hand column of Fig.~\ref{fig:m75_ray_evol}). 
            It is also possible to see that this shell moves at $u_r\approx10^9~\text{cm}~\text{s}^{-1}$, which is roughly one third of the speed of the wind. 
            Additionally, both the wind and stream are completely transparent as the radiative flux is identical along both lines of sight, explaining the isotropy of the source at large scale.
            
            These two models summarise both extremes of the impact of the surrounding stellar stream on the TDE evolution in the case of the fiducial models. 
            On one hand, low $M_\text{h}$ sets the timescale of the event and keeps stellar stream on the orbital plane. 
            If the stream is confined on the orbital plane its density is maximised, which makes more difficult to push it outwards and for radiation to penetrate through it. 
            As a consequence, the evolution differs significantly along $\theta=0$ and $\theta=90^{\circ}$, and such differences may last for hundreds of days. 
            On the other hand, a higher black hole mass allows the stellar stream to precess out of the orbital plane creating a complex structure. 
            However, the mass in the structure is the same fraction of the stellar mass but distributed over a larger volume. 
            Therefore, the impact of the structure on the wind and radiation of the event affects a larger solid angle but only at the early times of the evolution due to the lower density of the structure especially at larger scales. 
            As this work aims to characterise the impact of the precessed stream on the light curves we opted for not to include a detailed description of the radiation hydrodynamic evolution of the rest of the simulations. 
            However, the Supplementary Material of the article contains animations of the density and radiative flux maps of all the models analogous to maps shown on Figs.~\ref{fig:m60_maps} and~\ref{fig:m75_maps}. 
        \begin{figure*}
            \begin{subfigure}{0.495\linewidth}
                \caption{Model M60}
                \includegraphics[width=\linewidth]{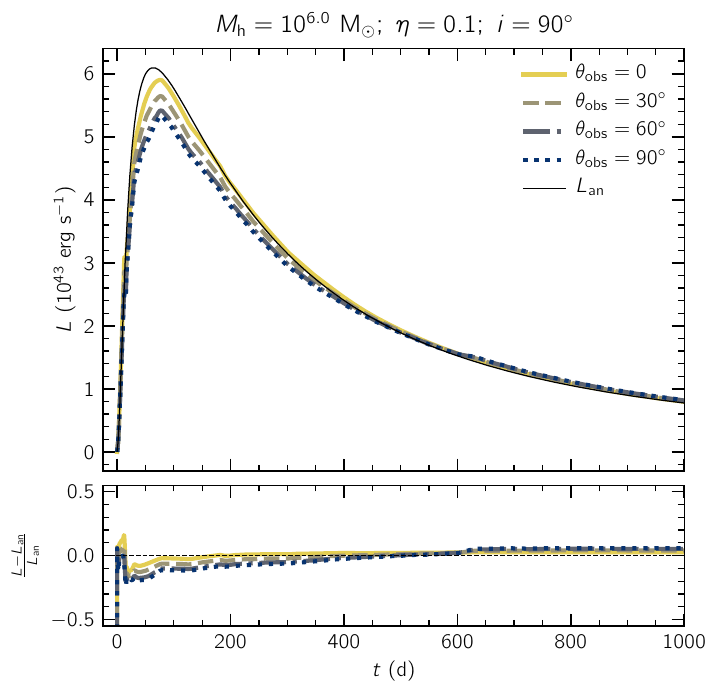}
                \label{fig:lc_m60}
            \end{subfigure}
            \begin{subfigure}{0.495\linewidth}
                \caption{Model M65}
                \includegraphics[width=\linewidth]{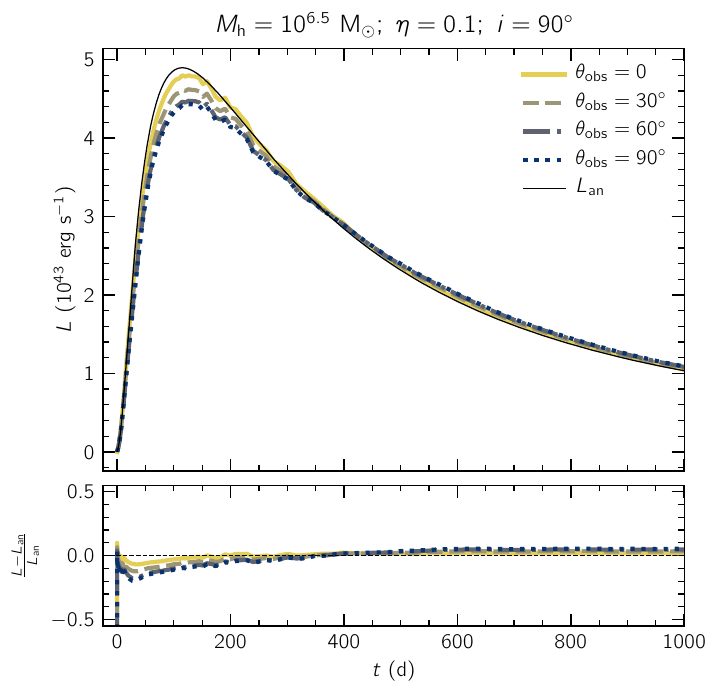}
                \label{fig:lc_m65}
            \end{subfigure}
            \begin{subfigure}{0.495\linewidth}
                \caption{Model M70}
                \includegraphics[width=\linewidth]{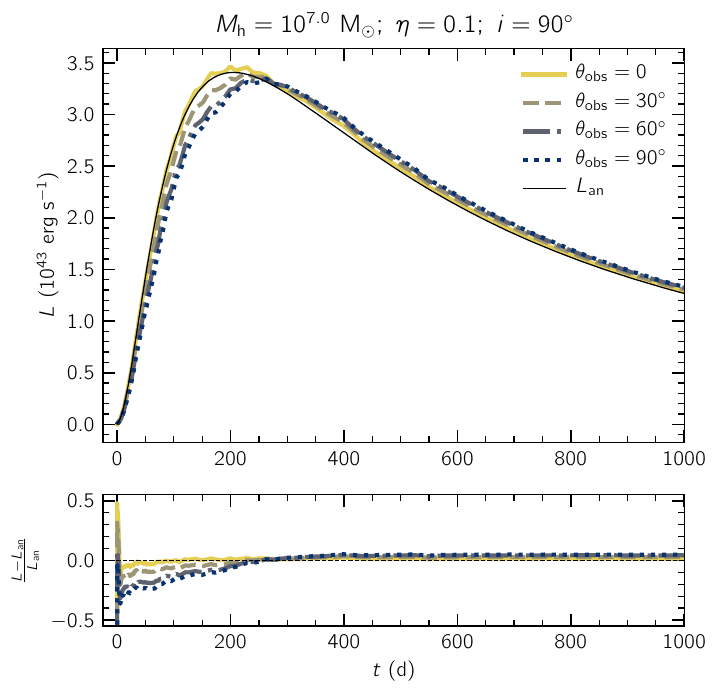}
                \label{fig:lc_m70}
            \end{subfigure}
            \begin{subfigure}{0.495\linewidth}
                \caption{Model M75}
                \includegraphics[width=\linewidth]{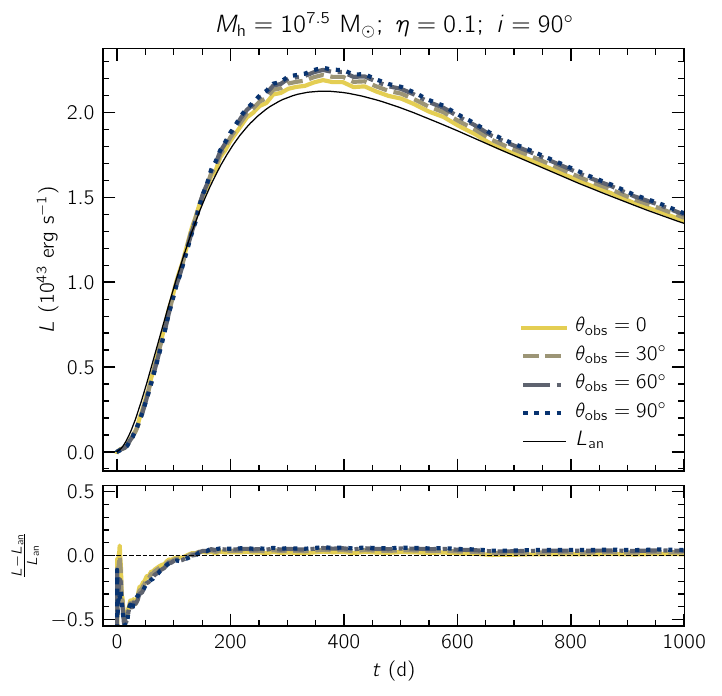}
                \label{fig:lc_m75}
            \end{subfigure}
            \caption{
            Simulated light curves computed from models M60 (\ref{fig:lc_m60}), M65 (\ref{fig:lc_m65}), M70 (\ref{fig:lc_m70}), and M75 (\ref{fig:lc_m75}). 
            Lines of sight $\theta_\text{obs}=0,~30^{\circ},~60^{\circ},~90^{\circ}$ are represented with solid yellow, dashed beige, dotted-dashed grey, and dotted blue lines, respectively. 
            The solid thin black line shows the semi-analytic model of the expected light curve of the event without surrounding medium following \protect\cite{piro2020}. 
            The lower part of each panel shows the relative difference between the simulated light curves and the semi-analytic calculation.}
            \label{fig:lcs}
        \end{figure*}
    \subsection{Light curve analysis}
    \label{sec:LCs}
        We proceed to study the light curves generated from each simulation.  
        To estimate the total radiated luminosity, first we map our two-dimensional model into three dimensions assuming azimuthal symmetry.  
        Then, we identify the photosphere location for each polar element integrating the optical depth along radial component to find $r_\text{ph}$ so that $\tau(r=r_\text{ph})=2/3$. 
        The observed luminosity for a given line of sight $\theta_\text{obs}$ was calculated integrated the radiated power over the whole photosphere weighing the contribution of each surface element by $\cos(\theta-\theta_\text{obs})$. 
        The result is the observed luminosity along a given line of sight as a function of time $L=L(\theta_\text{obs},t)$. 
        Additionally, we calculated the expected light curve for the wind and radiation injected using the semi-analytic approach developed by \cite{piro2020}, assuming that the opacity source is dominated by electron scattering. 
        The mass-loss rate and luminosity used were the same as in the simulations. 
        It is important remark that this calculation does not consider the presence of a surrounding medium as it assumes a spherically symmetric system. 
        The results are presented in the following subsections, describing the impact of the parameters of the models: black hole mass $M_\text{h}$, inclination $i$, accretion efficiency $\eta$, and mass fraction $f_\text{m}$.
        \begin{figure*}
            \includegraphics[width=0.475\linewidth]{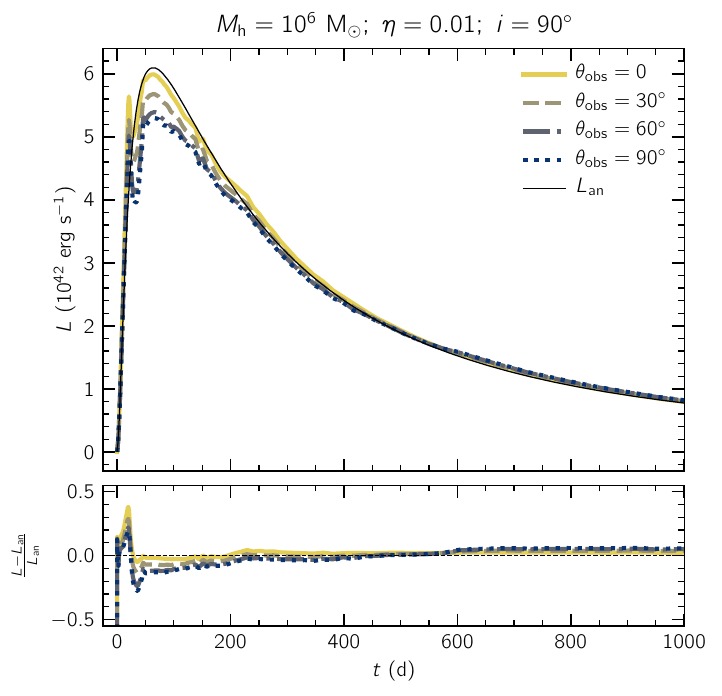}
            \includegraphics[width=0.475\linewidth]{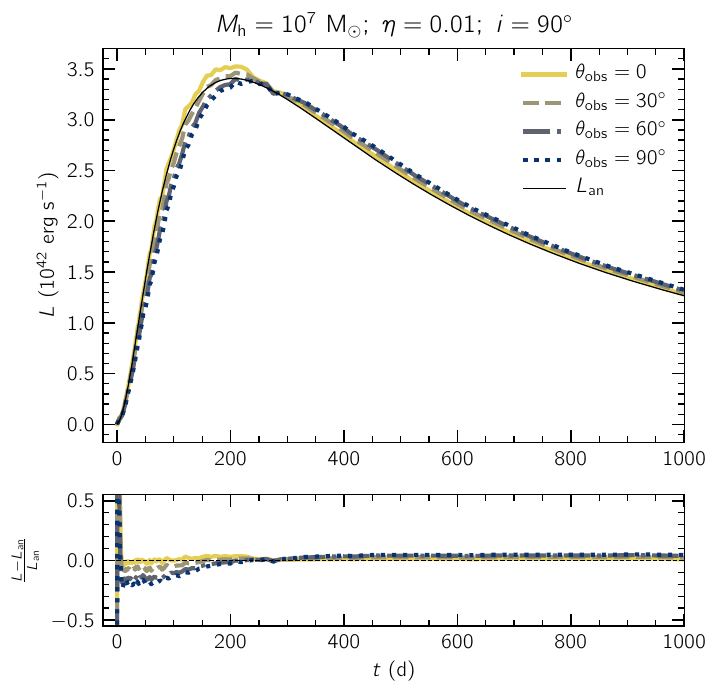}
            \caption{
            Simulated light curves computed from models M60-$\eta$01 (left-hand side) and M70-$\eta$01 (right-hand side). 
            Panels are analogous to panels in Fig.~\ref{fig:lcs}.
            }
            \label{fig:lcs_eta}
        \end{figure*}
        \subsubsection{Fiducial models: impact of black hole mass}
        \label{sec:lc_mass}
            The simulated light curves from the fiducial models M60, M65, M70, and M75 are shown in Fig.~\ref{fig:lcs}. 
            The observed luminosities along different lines of sight $\theta_\text{obs}=0$, $30^{\circ}$, $60^{\circ}$, $90^{\circ}$ are represented with solid yellow, dashed beige, dotted-dashed grey, and dotted blue lines, respectively. 
            The semi-analytic light curve $L_\text{an}$ is also included as a solid black thin line. 
            Panel~\ref{fig:lc_m60} contains the results computed from model M60, where it is possible to see the same qualitative shape of the light curves. 
            However, smaller values of the line-of-sight angle display a larger luminosity during the first half of the evolution ($\sim500~\text{d}$). 
            This is a consequence of the presence of the stellar stream only on the orbital plane. 
            Thus, both wind and radiation do not encounter resistance while traveling for lines of sight closer to the poles. 
            On the contrary, for an observer along the plane radiation do not travel directly, at least initially as it must diffuse through the stream. 
            Overall the differences are small ($\sim10~\text{per~cent}$) across different lines of sight. 
            Although there is an exception during the first tens of days of the event. 
            Here all light curves show a bump with an amplitude of at most $\sim20~\text{per~cent}$ for $\theta_\text{obs}$. 
            Such features are due to the wind hitting the stellar stream for the first time that heats up the dense material in the innermost edge of the stream. 
            This generates a transient bright flash that escapes more easily along smaller line-of-sight angles but it is quickly attenuated due to the wind becoming denser with time and trapping the radiation. 
            The main difference between the simulated and semi-analytic light curves is due to the assumption of electron scattering being the only source of opacity. 
            In the simulations, more opacity sources were considered that take into account, for instance, the absorption processes. 
            As a result, the radiation heats up the dense wind, and even manages to accelerate it to higher velocity than the launching speed (see Fig.~\ref{fig:m60_ray_evol}). 

            The light curves of model M65 show a very similar behaviour to model M60 (see Fig.~\ref{fig:lc_m65}). 
            In this case, the precession of the stream also acts only on the orbital plane, causing the same line-of-sight effects on the light curves as in the model M60. 
            However, the difference is that the outflow is less dense and, therefore the attenuation is smaller around the peak, which results into a better agreement with the semi-analytic model. 

            In the case of model M70, the light curves show their peaks at different times, being later for higher line-of-sight angle. 
            This is a result of the complex structure around the black hole due to precession out of the orbital plane that acts blocking and attenuating the radiation over a wider opening angle rather than solely along the orbital plane. 
            For $\theta_\text{obs}=0$, the light curves matches the analytic model due to the lower density of the wind, which is not enough to be accelerated by the radiation or to absorb significant energy from the wind. 
            The only exception occurs at early times ($t\lesssim10~\text{d}$), where the collision of the wind and the stream generates a bright transient feature. 
            Increasing $\theta_\text{obs}$, the radiation encounters directly the stream that is distributed over a larger polar angle (see Fig.~\ref{fig:ics}). 
            Once the light curve along $\theta_\text{obs}=90^{\circ}$ reaches its maximum ($t\approx250~\text{d}$), the source has become approximately isotropic. 

            For larger $M_\text{h}$, the impact of the surrounding medium on the light curve is confined only to the very early phase ($t\lesssim100~\text{d}$). 
            This is a direct consequence of the higher precession of the stream out of the orbital plane distributed the stellar material over a larger volume resulting in smaller density. 
            In Fig.~\ref{fig:lc_m75}, it can be seen that light curves are attenuated up to $50~\text{per}~\text{cent}$ relative to $L_\text{an}$, regardless of line of sight. 
            The fact that the stellar stream covers a wide solid angle only allows radiation to escape in direction of a small vicinity around the poles (see Fig.~\ref{fig:m75_maps}). 
            As a result, the observed luminosity differs significantly from the expected light curve until radiation manages to go through the stream. 
            It is important to bear in mind that a more massive black hole sets a larger length scale for the event, which translates into a less dense stream in general. 
            Thus, although its impact onto the observed light curves is 
            more significant it occurs on relatively shorter timescales.

            Through the analysis of varying the black hole there is one additional relevant aspect. 
            Notice that the location of the peak of these light curves does not seem to depend significantly on the observer angle with the exception of model M70. 
            This feature can be explained as a result of two competing effects caused by varying the black hole mass that are maximised in the case of M70. 
            On one hand, more massive black holes cause precession to be more significant outside of the orbital plane. 
            As a result, more lines of sight encounter the stellar stream, which in principle causes more light attenuation. 
            On the other hand, the density of the stream is smaller if more precession takes place as the same mass is distributed over a larger volume. 
            This decreases the attenuation of the light going through the stellar stream. 
            Thus, there is an optimal black hole mass vicinity where the attenuation is relevant and, at the same time affects a significant amount of lines of sight for long enough timescales. 
            That is why for more massive black holes we do not observe the extra attenuation around the peak but only during the rising to peak regime. 
            While for less massive black holes we do observe extra attenuation but it does not affect many lines of sight.
        \begin{figure*}
            \includegraphics[width=0.95\linewidth]{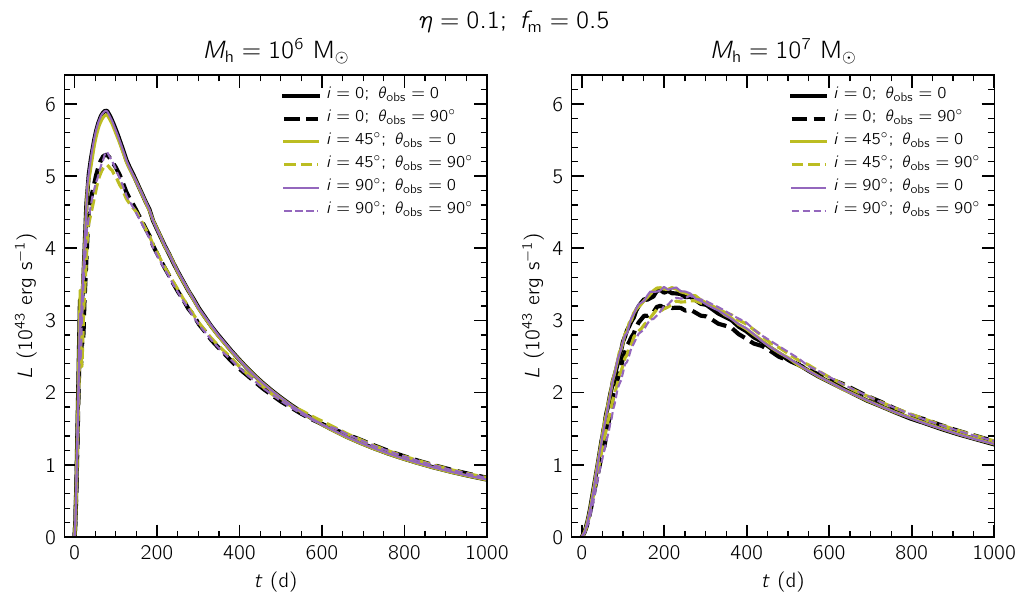} 
            \caption{
            Simulated light curves computed from models M60, M60-$i45$, M60-$i0$ (left-hand side) and M70, M70-$i45$, M70-$i0$ (right-hand side).
            Different inclination models $i=0,~45^{\circ},~90^{\circ}$ are shown as black, yellow, and purple lines, respectively.
            Additionally, the lines of sight $\theta_\text{obs}=0$ and $\theta_\text{obs}=90^{\circ}$ are represented with solid and dashed lines.}
            \label{fig:comparison_i}
        \end{figure*}
        \begin{figure*}
            \includegraphics[width=0.95\linewidth]{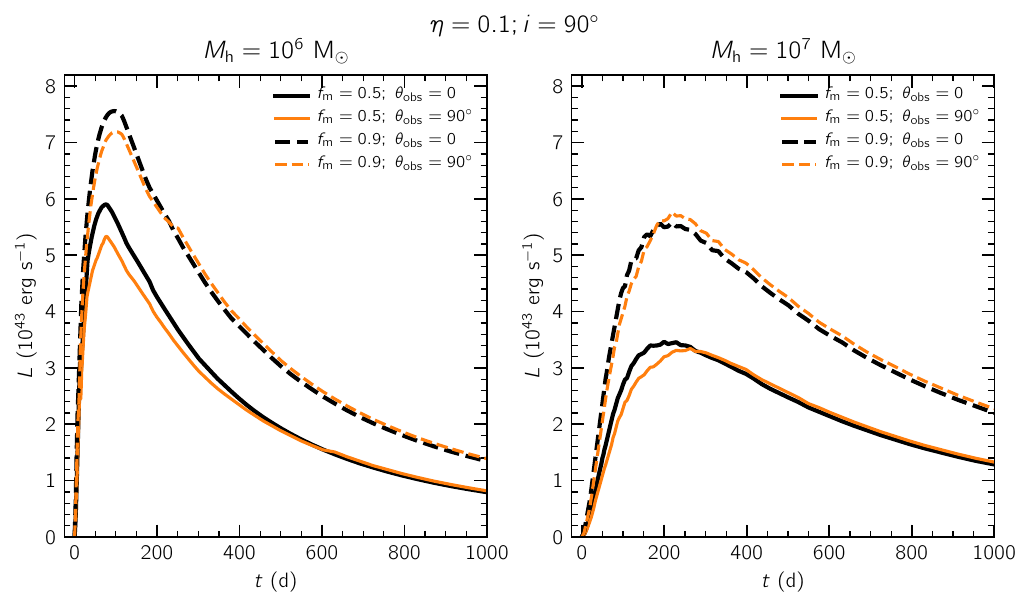}
            \caption{
            Simulated light curves computed from models M60, M60-$f9$ (left-hand side), and M70, M70-$f9$ (right-hand side). 
            Mass fractions $f_\text{m}=0.5$ and $f_\text{m}=0.9$ are shown as solid and dashed lines, respectively.
            Additionally, the lines of sight $\theta_\text{obs}=0$ and $\theta_\text{obs}=90^{\circ}$ are represented with black and orange lines.
            }
            \label{fig:lcs_fm}
        \end{figure*}
        \subsubsection{Low accretion efficiency: $\eta=0.01$}
        \label{sec:lc_eta}
            The accretion efficiency determines the accretion luminosity of the event, hence a smaller value decreases the luminosity of the event (see equation~\ref{eq:luminosity}). 
            As the kinetic energy of the outflow remains the same (see equation~\ref{eq:velmin}) its relative power increases, which results into a clear observational signature during the early phase of the event. 
            Fig.~\ref{fig:lcs_eta} shows the simulated light curves computed from the models with $\eta=0.01$: M60-$\eta01$ (left-hand side) and M70-$\eta01$ (right-hand side). 
            The panels are completely analogous to panels in Fig.~\ref{fig:lcs} in order to make comparisons more easily.
            
            Here, it can be observed that model M60-$\eta01$ displays a clear peak around $t\approx30~\text{d}$, deviating from the expected light curve (see left-hand side panel of Fig.~\ref{fig:lcs_eta}). 
            Notice that the peak amplitude is large with decreasing $\theta_\text{obs}$, reaching $\sim40~\text{per}~\text{cent}$ more relative to $L_\text{an}$. 
            This feature is caused by the collision of the outflow with the inner edge of the stellar stream. 
            The radiation generated can escape easily with lower line-of-sight angles due to the presence of the dense stream only on the orbital plane. 
            Immediately afterwards, the wind becomes dense trapping both the injected and generated radiation attenuating quickly the observed luminosity across all lines of sight but especially at $\theta_\text{obs}=90^{\circ}$.
            At later times, these light curves reproduce better the semi-analytic estimate, especially the one along $\theta_\text{obs}=0$. 
            This is a direct consequence of the lower luminosity, as radiation pressure is not strong enough to accelerate the wind further. 
            Thus, radiation exerts less work, causing most energy to remain in radiation and finally to propagate outwards.

            In the case of M70-$\eta01$ (right-hand side panel of Fig.~\ref{fig:lcs_eta}), a peak is also present at the beginning of the simulation but is qualitatively similar to the one observed in model M70 (see Fig.~\ref{fig:lc_m70}). 
            The higher relative amplitude of the peak is simply a consequence of the larger relative magnitude of the kinetic energy with respect to the injected luminosity, when compared to the fiducial case. 
            As a consequence, the light curves across all lines of sight are in some cases more luminous than $L_\text{an}$, especially at $\theta_\text{obs}=0$. 
            The rest of the evolution is analogous to the fiducial model.
        \subsubsection{Impact of inclination: $i=0,~45^{\circ},~90^{\circ}$}
        \label{sec:lc_inc}
            So far, we have explored models with extreme inclinations in order to maximise the precession out of the orbital plane. 
            Now, we proceed to analyse cases with moderate to no inclination, and compare them to the fiducial models. 
            Fig.~\ref{fig:comparison_i} shows the simulated light curves computed from models M60, M60-$i45$, M60-$i0$ (left-hand side) and M70, M70-$i45$, M70-$i0$ (right-hand side). 
            Light curves of models with $i=0,~45^{\circ},~90^{\circ}$ are shown with black, yellow, and purple lines, respectively; and were computed for lines of sight $\theta_\text{obs}=0$ (solid lines) and $\theta_\text{obs}=90^{\circ}$ (dashed lines).

            Models with $M_\text{h}=10^6~\text{M}_\odot$ (left-hand side panel of Fig.~\ref{fig:comparison_i}) do not show significant differences when comparing light curves for a given line-of-sight angle. 
            This behaviour is expected since inclination plays only a minor role in the outcome of the precession of the stellar stream for a black hole mass of $\sim10^6~\text{M}_\odot$. 
            However, small differences around the peaks of the light curves for $\theta_\text{obs}=90^{\circ}$ are observed. 
            We attribute this to different outcomes of the precession on the orbital plane that can result into different radial density distributions within the stream depending at the location where the stream collides with itself. 
            For instance, if the self-intersection of the stream occurs closer to the pericentre the density will be enhanced in this region, which can have impact on the earlier stages of the light curve and vice-versa. 
            Nonetheless, in general the impact is very small for a TDE involving a black hole of this mass. 

            In the case of models with $M_\text{h}=10^7~\text{M}_\odot$ (right-hand side panel of Fig.~\ref{fig:comparison_i}), only light curves observed along $\theta_\text{obs}=90^{\circ}$ are affected to different extent depending on the inclination.
            Specifically, increasing the inclination the stellar stream precesses out of the orbital plane covering a larger solid angle, causing extra attenuation of the escaping radiation at earlier times. 
            As a result, the peak of the light curve is shifted by $\sim50~\text{d}$ and the event remains slightly brighter afterwards for $\sim100~\text{d}$. 
            Nevertheless, if these events are observed along lines of sight close to the polar directions all light curves look indistinguishable. 
        \begin{figure*}
            \centering
            \includegraphics[width=0.95\textwidth]{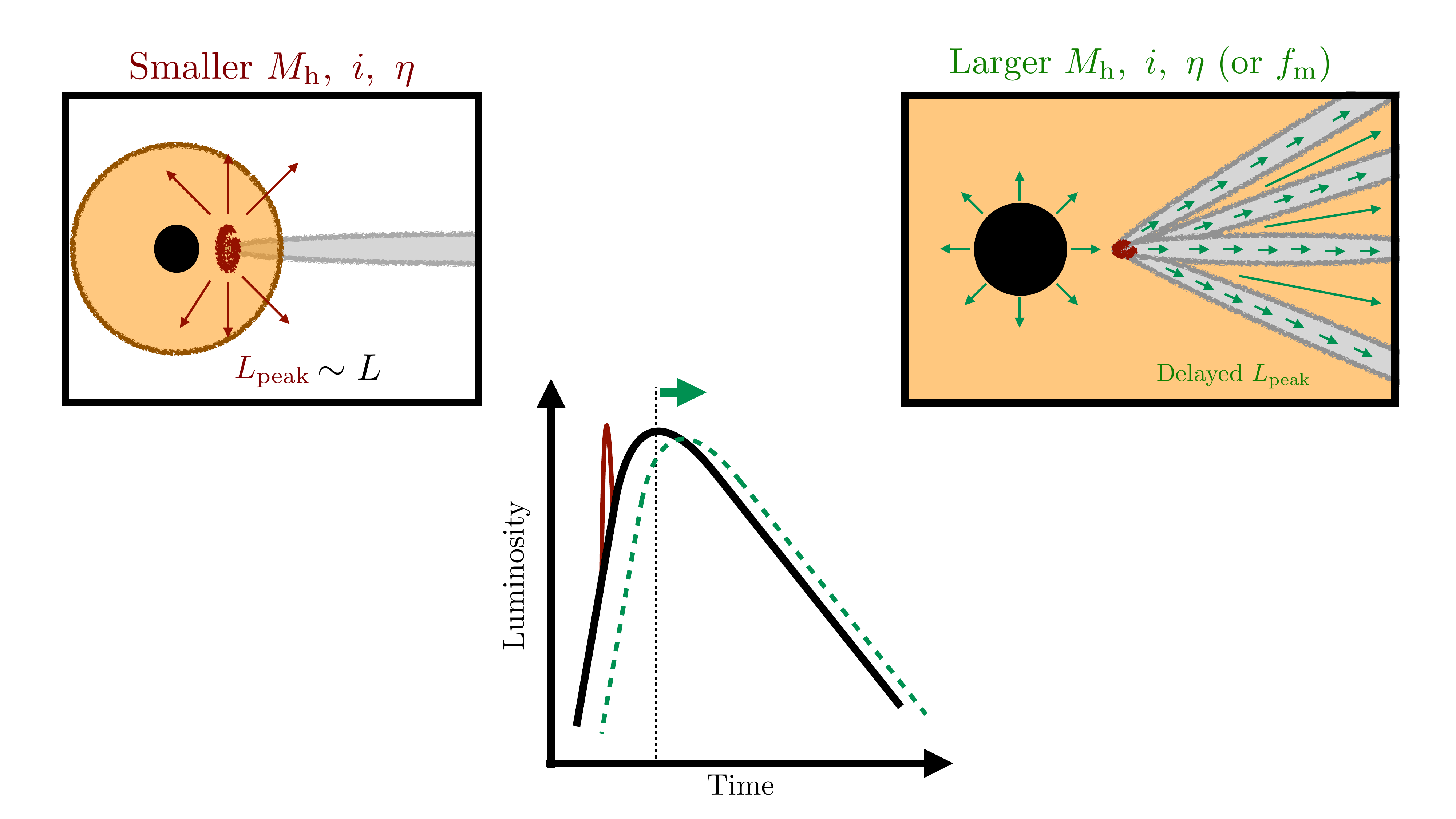}
            \caption{
            Schematic representation of both extremes of the impact of the surrounding stellar stream on TDE light curves.
            Models with $M_\text{h}\sim10^6~\text{M}_{\odot}$, and $\eta\sim0.01$ show light curves with an early sharp peak due to the interaction of the wind and the innermost part of the stream. 
            Models with $M_\text{h}\gtrsim10^7~\text{M}_{\odot}$, $i\sim~90^{\circ}$, and $\eta\sim0.1$ generate light curves whose peaks are delayed $50$-$100~\text{d}$ depending on the properties of the event and line-of-sight angle. 
            This is a result of the tidal stream covering a large solid angle that attenuates and blocks the radiation of the event, especially during the initial rising phase.}
            \label{fig:sketch}
        \end{figure*}
        \subsubsection{More accretion and massive wind, less massive stream: $f_\text{m}=0.9$}
        \label{sec:lc_fm} 
            So far we have assumed that half of the bound mass remains in the stellar stream and the other half falls onto the black hole. 
            In this section, we analyse the impact of this assumption through exploring the scenario, where only a small fraction of the bound material ($10~\text{per}~\text{cent}$) stays in the stellar stream and the rest is accreted or launched as a wind. 
            Fig.~\ref{fig:lcs_fm} shows the light curves computed from models M60 and M60-$f9$ (left-hand side panel), and M70 and M70-$f9$ (right-hand side panel). 
            In both panels, solid and dashed lines represent models with $f_\text{m}=0.5$ and $f_\text{m}=0.9$, respectively; while lines of sight are shown in colours black ($\theta_\text{obs}=0$) and orange ($\theta_\text{obs}=90^{\circ}$). 

            Before analysing the effect of $f_\text{m}$, it is necessary to bear in mind that the accretion luminosity is proportional to this parameter (see equation~\ref{eq:luminosity}) as well as the density of the outflow and, therefore the potential attenuation of the escaping radiation. 
            Thus, it is not obvious that the observed luminosity will indeed be higher in this case as the outflow is also denser, which may attenuate to a higher degree the injected luminosity.
            In the left-hand side panel of Fig.~\ref{fig:lcs_fm}, it can be seen that indeed models with $M_\text{h}=10^6~\text{M}_{\odot}$ and $f_\text{m}=0.9$ display larger observed luminosities with respect to the fiducial scenario. 
            This means that the net effect of increasing $f_\text{m}$ makes the luminosity brighter despite the denser wind. 
            Furthermore, the less dense mass in the stream decreases the deviations between observing the event either along the pole or orbital plane as expected. 
            In the right-hand side panel of Fig.~\ref{fig:lcs_fm}, 
            it is observed that a larger value of $f_\text{m}$ also produces a brighter event but the effect of the surrounding stream is different with respect to the fiducial case. 
            In this case, the stellar stream blocks part of the radiation of the event during the initial rise when observed along $\theta_\text{obs}=90^{\circ}$ in the same way than in the fiducial model. 
            Nevertheless, the behaviour of the light curves around the peak differ, as the light curve along $\theta_\text{obs}=90^{\circ}$ reaches a higher maximum $\sim50~\text{d}$ later than the peak along $\theta_\text{obs}=0$. 
            Less mass in the stream reduces the optical depth along the lines of sight that cross it, as a result the radiation generated in the collision of the outflow and the innermost part of the stream can diffuse faster with respect to the fiducial case. 
            The rest of the evolution of the light curve is analogous along both lines of sight.
            \\

            In summary, the surrounding stellar stellar stream can modify two aspects of the TDE light curves: i) the absence or presence of an extra early peak, and ii) the attenuation excess during the initial rise that could result into a delay of the expected peak in $50$-$100~\text{d}$. 
            Fig.~\ref{fig:sketch} shows an schematic representation of these two extremes, their impact on the light curves, and the required conditions for these to take place.
            Events involving less massive black holes ($M_\text{h}\sim10^6~\text{M}_{\odot}$) and low accretion efficiency ($\eta\sim0.01$) maximise the presence of the extra early peak. 
            This peak is powered by the collision of the outflow with the innermost side of the stellar stream, and its emission is more easily observed due to the higher relative power with respect to the TDE itself with low accretion efficiency.
            The amplitude of the peak is larger when observed at lower line-of-sight angle ($\theta_\text{obs}=0$), as larger angles will encounter part of the stream that would block the emission, especially during the initial phase. 
            On the other hand, events with more massive black holes ($M_\text{h}\gtrsim10^7~\text{M}_{\odot}$), high accretion efficiency ($\eta\sim0.1$), and non-zero inclination increase the effect of the early attenuation and peak delay. 
            The impact is maximum when observed along $\theta_\text{obs}=90^{\circ}$, decreases for smaller angles, and disappears along $\theta_\text{obs}=0$. 
            An important caveat is that in the case of high mass black hole, the  conclusion holds only if they are rapidly rotating ($a_\text{h}=0.9$). 
            This feature is key to produce significant precession of the stream out of the orbital plane that is the responsible for spreading it over a larger solid angle, producing the attenuation in the early phase and the delayed maximum in the light curve. 
            TDEs in black holes rotating more slowly will also show these signatures in their light curves but to less extent. 
\section{Discussion}
\label{sec:discussion}
    In this section, we discuss implications on observations, and limitations of the model and approach. 
    \subsection{Observational implications}
        Our simulations predict two potentially observable signatures in light curves as a result of the interaction between the TDE luminosity and wind with the tidal stream. 
        On one hand, there could be an early peak that would be more prominent in the case of low accretion efficiency and relatively low-mass black holes. 
        On the other hand, the rise to peak of the light curve could be attenuated causing a delay in reaching its maximum, provided that both the inclination and black hole mass are large. 
        Hence, it is sensible to wonder whether or not such features could be recognised in observed light curves. 

        The predicted early peak seen in the synthetic light curves has an amplitude comparable to the main maximum, and has a short duration $<30~\text{d}$ (see left-hand side of Fig.~\ref{fig:lcs_eta}). 
        To date, there are have been cases of light curves that show a re-brightening after the first peak \citep[e.g.][]{dong2015,leloudas2016,godoy2017}. 
        However, \cite{godoy2017} did not report any sign of interaction between the ejecta with the surrounding medium, which is typically inferred from the presence of narrow emission lines. 
        Additionally, they reported that the re-brightening occurred in ultraviolet bands, and not in visible bands. 
        Thus, it is hard to relate such a feature with the prediction of our model. 
        Nevertheless, this fact does not necessarily rule out this scenario, as a high accretion effiency ($\eta\sim0.1$) is enough to bury this early peak under the intrinsic luminosity of the event. 

        The attenuation and delay on reaching the maximum brightness are effects that may introduce more uncertainties into the analysis of light curves. 
        We have shown that the amount of light obscuration and its duration depend largely on the black hole mass, and varies with line-of-sight angle.
        Although this is a clear prediction from the model, it might be difficult to identify it in observed light curves since a delay in reaching the maximum might also be attributed to other effects such as the relativistic precession itself or the disc formation. 
        Hence, in practice this will add more uncertainty in light curve modelling rather than an easily recognisable feature caused by the wind-stream interaction. 

    \subsection{Limitations}
        \subsubsection{Two-dimensional approach}
            The simulations were performed by solving the two-dimensional radiation hydrodynamic equations in spherical coordinates, assuming azimuthal symmetry. 
            This approach was chosen in order to be able to explore a wide range of parameters of the problem without a high computational cost. 
            Sampling the three-dimensional tidal stream into a two-dimensional domain has two main effects. 
            First, it dilutes the mass in the stream by a factor $2\pi$, which reduces the density and may affect the optical depth of the stream. 
            Second, it increases the solid angle covered by the stream in the sky of the black hole, so that more line-of-sight angles will hit the stream when observing the event. 
            In reality, therefore the stream should be denser and cover a smaller solid angle in comparison with our simulations. 
            As a result, we expect that three-dimensional models would show more pronouced features in their light curves due to the interaction between the wind and the tidal stream but such effects may have a stronger dependence on the line-of-sight angle.  
            To properly quantify this issue, three-dimensional simulations must be developed, which we leave to a future work.
        \subsubsection{Grey approximation}
            To date, most TDEs have been detected in optical wavelengths, some of these show X-ray emission, and others have been detected only in X-ray \citep[e.g.][]{gezari2021}. 
            Unfortunately, it is not clear yet how and where the emission is generated as there is discrepancy on the inferred radii at which the radiation is produced. 
            Our models were carried out under the grey approximation, i.e. only frequency-integrated radiation. 
            On one side, this choice is based on simplicity due to the high computational cost and complexity of performing multi-group radiation hydrodynamics. 
            Additionally, since our model is agnostic about how the radiation is generated we decided not to investigate this aspect further, as our main goal was set to quantify the impact of the wind and luminosity with the tidal stream. 
            In this context, our result should be interpreted as guidelines to investigate further aspects of this scenario namely the radiation generated at different locations from the event or even the impact on the multi-group emission.
\section{Conclusions}
    We have investigated the effect of relativistic precession on TDE light curves. 
    To this end, we developed a set of moving-mesh radiation-hydrodynamic simulations of the interaction between TDE luminosity and disc wind with the leftover of the tidal stream wrapped around the black hole. 
    The model assumes that both luminosity and wind follow the fallback rate of a polytrope.  
    We investigated models involving rapidly spinning ($a_\text{h}\sim0.9$) super-massive black holes with masses in the range $10^{6.0}$-$10^{7.5}~\text{M}_{\odot}$. 
    Each model was simulated for a duration of at least $1000~\text{d}$.
    The simulations show that in all cases the wind structure is affected by the presence of the tidal stream, being forced into a bipolar shape rather than retaining its original spherically symmetric shape with which it is launched. 
    The opening angle of the outflow depends on the ability of the tidal stream to precess out of the orbital plane, and is therefore smaller for massive black holes with high inclination, and larger for black holes with smaller masses or small inclination. 
    
    We estimated light curves for different lines of sight for each simulated model. 
    This analysis allowed to us to quantify the impact of the accretion efficiency, inclination, black hole mass, and the amount of mass assumed to be in the surrounding stream. 
    From this, we were able to identify two cases based on the properties of the event and the signatures imprinted in their light curves (see Fig.~\ref{fig:sketch}): 
    $\text{i)}$ events involving black holes with $M_\text{h}\sim10^6~M_{\odot}$ and low accretion efficiency ($\eta\sim0.01$), and 
    $\text{ii)}$ events with more massive black holes with $M_\text{h}\gtrsim10^7~M_{\odot}$, large inclination ($i\sim90^{\circ}$), and high accretion efficiency ($\eta\sim0.1$). 
    The former type of events have light curves that show a fast and sharp peak before the maximum of the fallback rate is reached. 
    This feature is produced by to the interaction of the disc wind and the tidal stream innermost edge. 
    Events with higher accretion efficiencies also show this interaction and feature but the lower power relative to the luminosity of the event hides it below the main signal. 
    The later class of events show light curves that are attenuated during the initial rise due to the presence of the precessed tidal stream. 
    In this case, the effect is more pronounced because a more massive black hole and a high inclination produce a precession out of the orbital plane of the stream, which spreads it over a larger solid angle. 
    As a result, a wider range of line-of-sight angles are affected by the stream blocking the radiation, generating a delay of $50$-$100~\text{d}$ relative to the expected time of the light curve peak. 

    Although our models predict clear signatures on TDE light curves we conclude that such features may be present in the observed light curves but it might not be possible to identify them unambiguously. 
    The results of this work should be interpreted as a first approach for constraining the effect of relativistic precession, as we have worked under the grey and azimuthal symmetry assumptions. 
    Our results have allowed us to identify the potential observable features under the most relevant cases, where precession can have the most significant impact on the light curves. 
    With this in hand, we can proceed further to investigate such models with a more realistic, albeit more computationally expensive setup, such as a three-dimensional approach and/or multi-group radiation-hydrodynamics. 
    
\label{sec:conclusions}
\section*{Acknowledgements}
    We would like to thank the anonymous referee for useful comments and suggestions that contributed to improve this article.
    The research of DC and OP has been supported by Horizon 2020 ERC Starting Grant `Cat-In-hAT' (grant agreement no. 803158). 
    Since 03/2023, DC was funded by the Deutsche Forschungsgemeinschaft (DFG, German Research Foundation) under Germany’s Excellence Strategy - EXC 2121 - ``Quantum Universe” - 390833306.
    This work was supported by the Ministry of Education, Youth and Sports of the Czech Republic through the e-INFRA CZ (ID:90140 and ID:90254).
    Part of this work was developed at the Aspen Center for Physics, which is supported by National Science Foundation grant PHY-1607611. 
    In addition, DC acknowledges the kind hospitality of the Center for Computational Astrophysics, where part of this project was conducted.
    This work made use of \textsc{python} libraries \textsc{numpy} \citep{harris2020} and \textsc{matplotlib} \citep{hunter2007}, as well as of the NASA’s Astrophysics Data System.
\section*{Data Availability}
    The output files from our simulations will be shared on reasonable request to the corresponding author.
\bibliographystyle{mnras}
\bibliography{tde} 
\appendix
\section{Impact of spin parameter on the initial conditions}
\label{sec:app}
\begin{figure*}
        \centering
	\includegraphics[width=0.95\linewidth]{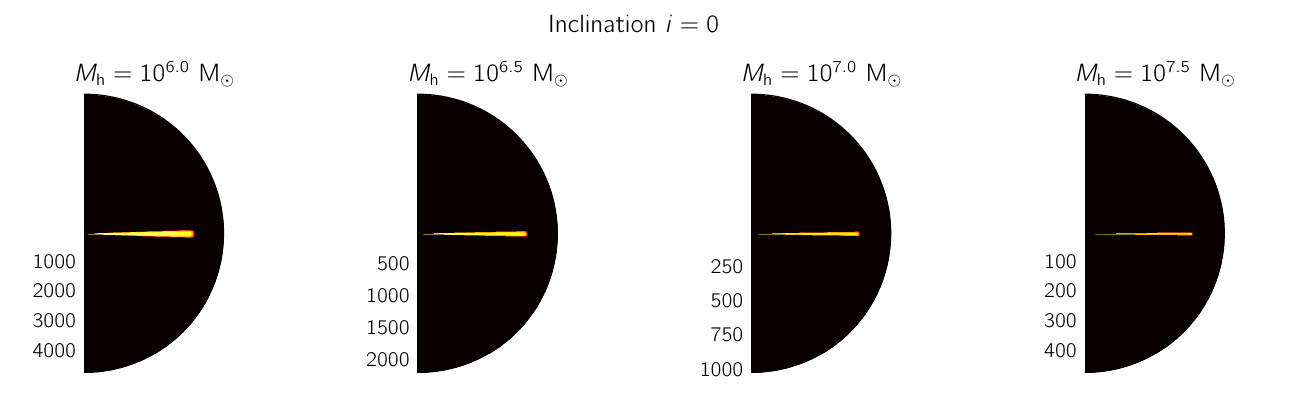}
	\includegraphics[width=0.95\linewidth]{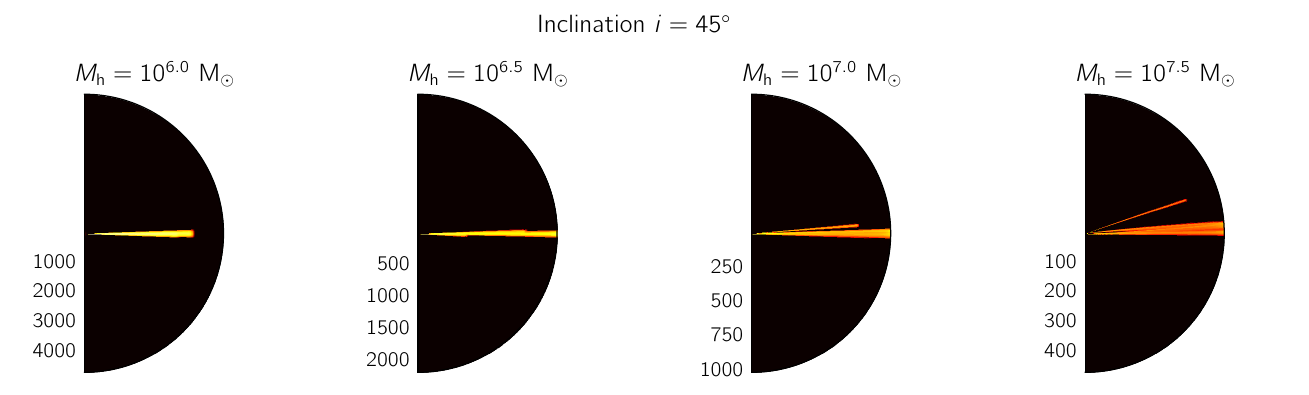}
        \includegraphics[width=0.95\linewidth]{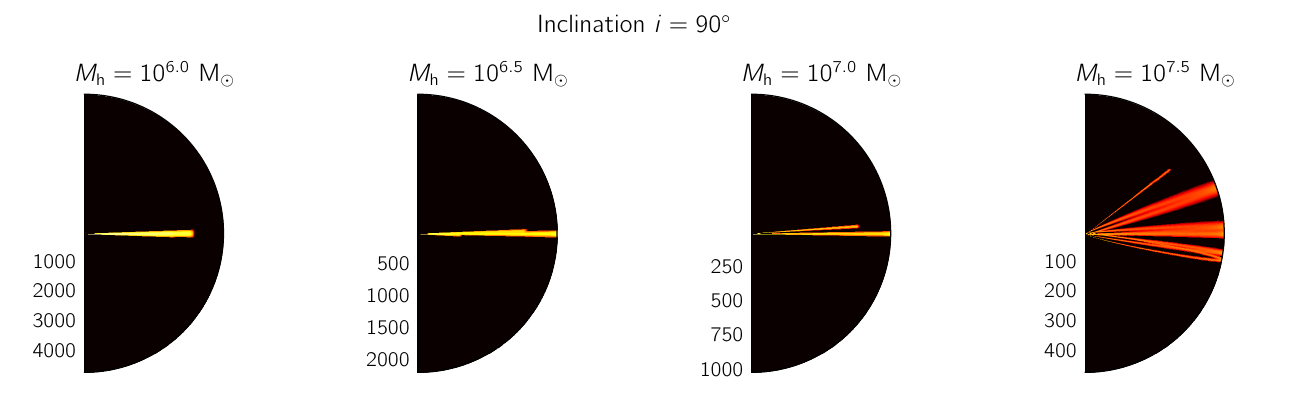}
        \includegraphics[width=0.5\linewidth]{plots/ICs_colorbar.pdf}
        \caption{
        Analogous to Fig.~\ref{fig:ics} but for $a_\text{h}=0.5$.
        Two-dimensional $(r,\theta)$ density maps of the precessing TDE model for varying inclination $i$ and black hole mass $M_\text{h}$. 
        Upper, central, and lower rows show models with inclination $i=0$, $45^{\circ}$, $90^{\circ}$, respectively. 
        Columns contain panels with a given black hole mass, from left- to right-hand side $M_\text{h}=10^{6.0}$, $10^{6.5}$, $10^{7.0}$, $10^{7.5}~\text{M}_{\odot}$. 
        Radial spatial scales are shown in units of Schwarzschild radii according to their black hole mass, i.e $R_\text{Sch}=2GM_{\rm h}/c^2$.
        }
        \label{fig:appendix_ics}
    \end{figure*}

    The black hole spin $a_\text{h}$ indeed plays an important role in shaping the precessed stream around the black hole. 
    If the black hole is not spinning fast enough ($a_\text{h}\sim0.5$) the precession out of the orbital plane occurs only for extreme values of both black hole mass ($M_\text{h}\sim10^{7.5}~\text{M}_{\odot}$) and inclination ($i\sim90^{\circ}$). 
    Fig.~\ref{fig:appendix_ics} illustrates this fact as it shows the precessed stellar stream for cases with $a_\text{h}=0.5$, different inclination $i=0,45^{\circ},90^{\circ}$ along each row (top to bottom), and different black hole mass $M_\text{h}=10^6,10^{6.5},10^7,10^{7.5}~\text{M}_{\odot}$ along each column (left- to right-hand side). 
    Notice that in this case most structures are aligned with the orbital plane with the exception of the models with the highest mass and inclination. 

\bsp	
\label{lastpage}
\end{document}